\documentclass [12pt,twoside]{article}           
\usepackage[left,modulo]{lineno}
\usepackage{setspace}

\countdef\arxiv=201

\ifnum\arxiv=0 \input cpreb.tex \fi

\ifnum\arxiv=1

\usepackage{epsfig,times,lscape}
\usepackage[usenames]{color}

    \ifnum\count102=1

    \fi

    \ifnum\count102=2
\fi

\newcommand{\nc}{\newcommand}

     \ifnum\count101=1
\nc{\qI}[1]{\section{{#1}}}
\nc{\qA}[1]{\subsection{{#1}}}
\nc{\qun}[1]{\subsubsection{{#1}}}
\nc{\qa}[1]{\paragraph{{#1}}}

\def\qpar{\vskip 2mm plus 0.2mm minus 0.2mm}
\def\qL{\hfill \break}
     \fi 

      \ifnum\count101=2
 \nc{\qI}[1]{\parindent=0mm \vskip 8mm 
{\centerline{\LARGE \color{red}#1}}\vskip 3mm}
\nc{\qA}[1]{\vskip 2.5mm \noindent 
{{\bf\large\color{blue}  #1}} \vskip 1mm \parindent=0mm}
 \nc{\qun}[1]{\vskip 1mm \noindent {\sl #1 }\quad }

\def\qL{\hfill \break}
\def\qpar{\vskip 2mm plus 0.2mm minus 0.2mm}

      \fi

\def\qth{\vrule height 12pt depth 0pt width 0pt}
\def\qtb{\vrule height 0pt depth 5pt width 0pt}

\nc{\qfoot}[1]{\footnote{{#1}}}

      \ifnum\count101=1
\def\qbu{\hfill \par \hskip 6mm $ \bullet $ \hskip 2mm}
\def\qee#1{\hfill \par \hskip 6mm (#1) \hskip 2 mm}
      \fi
      \ifnum\count101=2
\def\qbu{\hfill \par \hskip 4mm $ \bullet $ \hskip 2mm}
\def\qee#1{\hfill \par \hskip 4mm (#1) \hskip 2 mm}
      \fi

\def\qparr{ \vskip 1.0mm plus 0.2mm minus 0.2mm \hangindent=10mm
\hangafter=1}

     \ifnum\count101=1 
 \def\qdec#1{\parindent=0mm\par {\leftskip=2cm {#1} \par}}
     \fi
     \ifnum\count101=2
  \def\qdec#1{\parindent=0mm \par {\leftskip=1cm {#1} \par}}
  
  \def\qcitb#1{\noindent \hbox to 102mm{\hfill \small #1} \vskip 1mm}
      \fi

 \def\qpages#1{\count102=0{\loop\advance\count102 by 1
 \null \vfill\eject \ifnum\count102<#1 \repeat}}

\def\qn#1{\eqno \hbox{(#1)}}

\def\qth{\vrule height 12pt depth 0pt width 0pt}
\def\qtb{\vrule height 0pt depth 5pt width 0pt}

\def\qv{\vskip 0.1mm plus 0.05mm minus 0.05mm}
\def\qhu{\hskip 0.6mm}
\def\qhv{\hskip 3mm}

\def\qhw{\hskip 1.5mm}
\def\qleg#1#2#3{\noindent {\bf \small #1\qhw}{\small #2\qhw}{\it \small #3}\qv }
\newcommand{\promille}{%
  \relax\ifmmode\promillezeichen
        \else\leavevmode\(\mathsurround=0pt\promillezeichen\)\fi}
\newcommand{\promillezeichen}{%
  \kern-.05em%
  \raise.5ex\hbox{\the\scriptfont0 0}%
  \kern-.15em/\kern-.15em%
  \lower.25ex\hbox{\the\scriptfont0 00}}

\fi

\begin{document}
\thispagestyle{empty}



\markboth{{\sl \hfill  \hfill \protect\phantom{3}}}
        {{\protect\phantom{3}\sl \hfill  \hfill}}

\color{yellow} 
\hrule height 10mm depth 10mm width 170mm 
\color{black}

 \vskip -17mm   

\centerline{\bf \Large Coupling between death spikes
and birth troughs.}
\vskip 5mm
\centerline{\bf \Large Part 1: Evidence}
\vskip 10mm

\centerline{\large 
Peter Richmond$ ^1 $ and Bertrand M. Roehner$ ^2 $
}

\vskip 10mm
\large

%
{\bf Abstract}\qL
In the wake of the influenza pandemic of 1889-1890 Jacques Bertillon,
a pioneer of medical statistics,
noticed that
after the massive death spike there was a dip in birth numbers around
9 months later
which was significantly larger than that which could be explained by
the population
change as a result of excess deaths. In addition it can be 
noticed that this dip was followed by
a birth rebound a few months later. However having made this
observation, Bertillon
did not explore it further. Since that time the phenomenon was
not revisited in spite of the fact that in the meanwhile there
have been several new cases
of massive death spikes. The aim here is to analyze these new
cases to get
a better understanding of this death-birth coupling phenomenon.
The largest death spikes occurred in the wake of more recent influenza
pandemics in
1918 and 1920, others were triggered by the 1923 earthquakes in Tokyo
and the Twin
Tower attack on September 11, 2001. We shall see that the first of
these events indeed produced
an extra dip in births whereas the 9/11 event did not.
This disparity highlights the pivotal role of collateral sufferers.
In the last section it is shown how the present coupling leads
to predictions; it can explain 
in a unified way
effects which so far have been studied separately, 
as for instance the impact on birth rates of heat waves. 
Thus, it appears that
behind the random appearance of birth rate fluctuations there
are in fact hidden explanatory factors.
                              
\vskip 10mm
\centerline{\it \small Version of 14 January 2018}
\vskip 5mm

{\small Key-words: death spike, birth rate, influenza, pandemic,
earthquake, 9/11} 

\vskip 5mm

{\normalsize
1: School of Physics, Trinity College Dublin, Ireland.
Email: peter\_richmond@ymail.com \qL
2: Institute for Theoretical and High Energy Physics (LPTHE),
University Pierre and Marie Curie, Paris, France. 
Email: roehner@lpthe.jussieu.fr
}

\vfill\eject

\qI{Introduction}

This paper is about a remarkable case of birth and death
fluctuations which, apart from its own interest, may give new
insight in the more general problem of vital rate fluctuations.
In this respect it may be useful to start with the following
observation.

\qA{Two alternative views of birth fluctuations}

The number of births in a given time interval $ \theta $ 
can be written: $ b(t)=\lambda (t)P(t) $ where $ \lambda (t) $
is the crude death rate and $ P(t) $ the size of the population. 
Depending on whether
$ \theta $ is one week, one month or one year, $ \lambda $ will
be the weekly, monthly or annual rate.
\qpar
 
For the
fluctuations of $ b(t) $ the previous formula leads us to distinguish
two cases:
\qee{1} The rate is fixed $ \lambda(t)=\lambda $. In this case
the changes of $ b(t) $ reflect the changes of the
population.
\qee{2} The population is constant $ P(t)=P $. In this case the 
fluctuations of $ b(t) $ reflect the fluctuations of the
birth rate $ \lambda (t) $. 
\qpar

Conceptually, there is a great difference between these cases for
in the first case there is nothing to explain whereas the second
case raises many questions. What we mean by ``nothing to explain''
is that, as the population increases or decreases
regularly the only question that one may possibly
raise is why the birth rate is set at its given value $ \lambda $.\qL
On the contrary, the fluctuations of the rate raises
numerous questions in the sense that $ \lambda (t) $ reflects
the behavior of the individuals and the conditions under which they
live. In other words, for any major change in $ \lambda (t) $
one will be prompted to ask to what change in real life it is
tied. 
\qpar

A quick comparison of the orders of magnitude of the two
types of fluctuations can be made as follows.
\qbu In the early 20th century European populations
were increasing at an annual rate of around 1\%
Under the assumption of a constant $ \lambda $
birth numbers will see the same annual change. Thus,
for {\it monthly} changes the rate will be 12 times smaller,
i.e. around 0.1\%.
\qbu In contrast, the fluctuations of actually observed
monthly birth numbers (or birth rates)
were about 4\%
\qfoot{The 4\% estimate is for normal conditions; exceptional
birth dips and rebounds can be much larger, for instance
$ \pm 30\% $ in the case of Fig. 1b.}%
.
This is 40 times larger than
the fluctuations under constant birth rate.
\qpar

However remarkable,
the case considered in this paper is possibly just one of several
similar processes leading to predictable
birth rate changes. The topic of short-term
fluctuations has so far attracted much 
less attention than the study of medium- or long-term changes.
For instance the demographic transition in developed countries
was brought about by
changes of vital rates over a time interval of several decades.
\qparr

The present study shares several important features with the 
research field that is concerned with the fluctuations of
sex ratio at birth, see James (1971,2010).
\qbu Both researches rely crucially on comparative analysis,
either in space across different countries or in time over
past centuries. 
\qbu Both investigations focus on short-term effects, for instance
the changes that occur in the months following an epidemic or a war,
see Polasek et al. (2005) and James (2009).
\qbu The respective key-variables, monthly birth rates on one hand
and sex ratio fluctuations on the other, can be used as markers,
in other words as measurement devices which give insight into
abnormal situations. For instance, James (2006) documents how
offspring sex ratios can reveal endocrine disruptions; similarly
mortality sex ratios can be used to explore anomalies of
the  immune system,
see Garenne (1994) and Garenne et al. (1998).

\qA{Are there ``hidden variables''?}

In any country the time series of monthly births display
substantial fluctuations.
There is usually a seasonal pattern which is
country dependent; in addition for the same month (say October)
in different years there are annual fluctuations of about the same
magnitude. It is customary to say that these are random
fluctuations but are they really random? In this respect
one can observe that by saying that a phenomenon is random
one gives up {\it ipso facto} all attempts to understand it.
It is probably for this kind of reason that Albert Einstein
supported the ``hidden variable'' interpretation of
quantum physics. In 1935, i.e. some 10 years after 
quantum mechanics was introduced, Einstein (1935) suggested
that the wave function description of quantum objects was
incomplete in the sense that it did not include some hidden
parameters. 

\qA{The Bertillon discovery}

In 1892 Jacques Bertillon, a pioneer of medical statistics and
one of the designers of the ``International Classification of Diseases'',
published an analysis of the influenza pandemic of November 
1889--February 1990 in which he 
showed that approximately 9 months after
the climax of the epidemic a temporary birth rate trough (of an amplitude
of about 20\%) was observed in all countries where 
the pandemic has had a substantial impact, particularly Austria, France, 
Germany or Italy. 
\qpar
While writing his paper Bertillon did not know that some 28 years later
there would be a massive influenza pandemic through
which his discovery could be tested. We will show below that 
it was indeed confirmed in all countries where the pandemic
has had an impact. 
Apparently, there has been no further studies of this phenomenon
ever since. 
Actually, even in Bertillon's paper the effect
is discussed fairly briefly. In particular its mechanism
remains to be uncovered. This is the purpose of the present paper.
\qpar

While in 1889-1990 and 1918-1919 the effect can be detected
very clearly, is it not natural to assume that it exists
also in less spectacular cases. This raises the following question.
In most countries the death rate presents a winter peak
in January or February.
According to the Bertillon coupling one would 
expect a trough of births 9 months later that
is to say in October or November. Is this indeed the case?
If not, in what respect do exceptional death surges
differ from regular winter surges?
This point will be discussed in the second one of the
two papers devoted to this question%
\qfoot{In this respect see also Sardon 2005.}%
.
\qpar

In an attempt to get a clearer idea of the mechanism of the
coupling effect
the present paper will address the following questions.
\qee{1} First, we will review and expand the analysis presented in
Bertillon (1992). The question which comes immediately 
to mind is whether
the same effect can be observed in other pandemics.
\qee{2} If the answer to the previous question is affirmative
it raises another interrogation, namely is this effect 
restricted to pandemics or does it also
exist for other large-scale mortality shocks, for instance
famines, earthquakes and so on. Wars should not be included
in our study for in this case the separation between husbands and
wives interferes with the coupling that we wish to observe.
\qee{4} A possible mechanism that one can imagine is
through a transient impact 
on marriages. If the marriage rate is reduced
during the time of the epidemic, one would indeed expect
a reduction in birth numbers some 9 months later. 
Such an explanation can easily be
tested provided one can find {\it monthly} marriage data
(see below).

\qI{What is the {\it direct} incidence on births of mortality spikes?}

Before starting this investigation we need to answer an obvious
objection which can be stated as follows.
\qdec{Is it not natural that following a reduction 
in population one sees
a fall in the number of births? After this fall the birth numbers
will go up again as the
population resumes its ascending movement.}
\qpar

Qualitatively the argument seems satisfactory. If 
correct, the effect would become trivial. 
However, in what follows we show that quantitatively
the argument is not correct for it explains less than one tenth
of the observed birth changes.
\qpar

Before coming to more elaborate arguments one can make a simple
remark. What we see after 9 months is a fairly narrow dip.
On the contrary, a fall in population would produce a permanent
reduction, in other words not a dip but a Heaviside step.
It is true that because of the upward trend due to the overall
population growth after a while
the births would resume their ascending progression, however
the resulting shape would be a broad trough rather than a narrow
dip; this difference can be seen clearly in Fig. 1a versus 1b.
\qpar  

We present the reasoning in two forms. While straightforward,
the first one is
perhaps not very transparent; the second is more
theoretical but intuitively clearer.

\qA{Prediction of birth numbers under the assumption of a 
constant birth rate}

%
\begin{figure}[htb]
\centerline{\psfig{width=17cm,figure=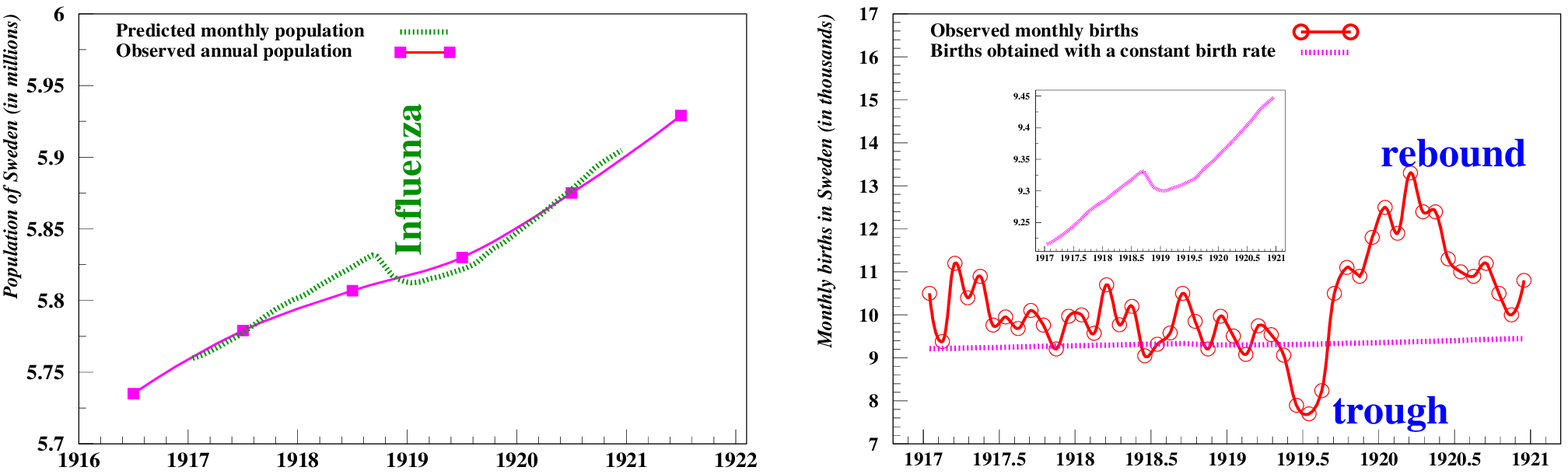}}
\qleg{Fig.\qhu 1a,b\qhv Comparison of the births produced
through population changes under constant birth rate on the
one hand and observed monthly fluctuations
of birth numbers on the other hand.}
{The data are for Sweden. 
Fig 1a shows that the population changes based on the summation
of monthly births and deaths approximates correctly the
observed annual population changes. The curve of birth numbers based
on a constant birth rate is naturally
parallel to the population curve; it is displayed in the inset
graph of Fig, 1b; the variation interval of births numbers 
(as shown on 
the vertical axis of the inset) is (9.20,9.45).
These variations are much smaller
than the actual fluctuations of birth numbers; 
as a result the broken
line of predicted birth numbers
looks almost completely flat when displayed on the same graph
as the actual births fluctuations.}  
{Source: Bunle (1954).}
\end{figure}

The first argument is very simple.\qL
The statistical records provide monthly birth and death numbers,
$ b_e(t) $ and $ d_e(t) $. Thus, starting from a known population
at initial time $ t_0 $ we can forecast the population $ P_f(t) $
in all subsequent months simply by summing up the monthly
population increases $ b_e(t)-d_e(t) $. It can be checked (Fig. 4a)
that $ P_f(t) $ is indeed consistent
with the observed population evolution $ P_e(t) $ (which amounts to
say that in this time interval
emigration and immigration do not play a great role).
Then, under the assumption of a constant birth rate $ \lambda $
the predicted monthly birth numbers will be: $ b_f(t)=\lambda P_f(t) $.
When these numbers are compared with the 
observed monthly birth numbers 
$ b_e(t) $ one sees a huge difference
in the magnitude of the fluctuations (Fig. 1b).
\qpar
To double check that there is no flaw in the data on which the graph
is based let us consider the start and end points. According to
Flora (1987, p.73):
\qbu In 1917 when the the mid-year population was 5,779 thousands 
a (constant)
annual birth rate of 20 per thousand gives 115,580 births for the
year, i.e. an average of 9,631 births per month.  
\qbu In 1920 when the mid-year population has increased to 5,875
the same birth rate gives 117,500 births, i.e. an
average of 9,791 per month. This represents a total 
predicted birth number increase of 1.67\% 
whereas the observed monthly birth numbers show
fluctuations of the order of 30\%.

\qA{Implication of a constant birth rate for the 
fluctuations of birth numbers due to population changes}

The following statement (explained in Appendix A) summarizes
the situation.
\qdec{{\bf Proposition}\quad Under the assumption of a 
constant monthly birth rate $ \lambda  $
the fall in births resulting from a monthly death excess $ e $ is
given by: $ \Delta b=\lambda e $;  for European countries in the
early 20th century $\lambda \simeq 2\%/12 $ which leads to:
$ \Delta b \simeq e/600 $.}  
\qpar

The proposition has a fairly clear intuitive interpretation.
To make things simple let us assume that the 
death-excess variable $ e $ takes place in one 
month, as is for instance the case for earthquakes.
In a virtual world where everybody conceives once 
every month, for each missing person there would be a missing
baby 9 months later. Under this assumption the birth
trough would be of same magnitude as the death spike.
However we know that actually the probability to conceive in a given
month is $ \lambda/12=0.0017 $ where $ \lambda $ is the
annual birth rate. In other words in 10,000 persons
only 17 may conceive in the relevant month.
Thus, this direct effect contributes very little
to the birth trough.

\qI{From conception to birth}

\qA{The role of collateral sufferers}

The birth rate reduction can 
be attributed to one of the factors mentioned in
Fig. 2a. The objective of the present paper is to
see more clearly which one of these factors plays a leading
role. What real-life mechanisms can one think of that may explain the
temporary birth rate fall? 

%
\begin{figure}[htb]
\centerline{\psfig{width=12cm,figure=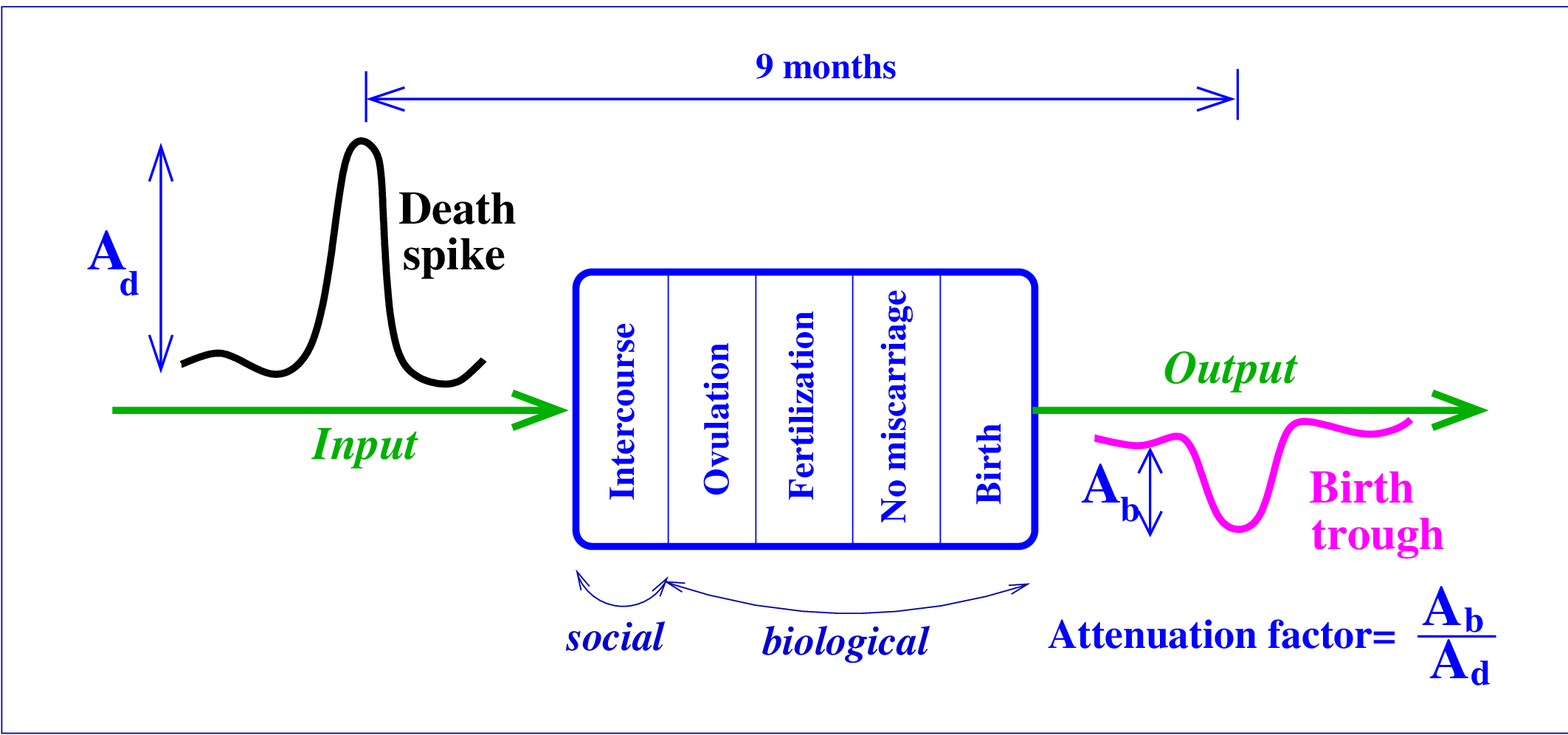}}
\qleg{Fig.\qhu 2a\qhv Representation of the
Bertillon coupling effect in the form of 
an input/output system.}
{If one interprets the 
death spike as an impulse function,
the birth trough can be seen as the impulse response of the
system. An important point is that, as explained in the
text, the loss of lives during the death spike cannot
by itself account for the observed birth trough.
Its direct effect is at least 10 times too small
which means that there must also be an indirect effect;
it is the purpose of the paper to identify it
more closely.}
{}
\end{figure}
\qpar

In Fig. 2a there is a distinction between social and biological
factors. It is by comparing different case studies that we
came to the conclusion that social factors play a key-role.
How?\qL
As famine and influenza have biological effects, 
at first sight the resulting
birth reduction might also be attributed to such effects. 
On the contrary, birth reductions after earthquakes can hardly 
by accounted for by biological factors. For that reason
the case of the Tokyo earthquake 
of 1923 played a key role in our understanding. It convinced 
us that the suffering of the many people who are affected but 
do not die was of central importance.
\qpar

In examining successive case-studies it will also be seen that the
magnitude of the birth reduction is compatible with
the existence of a group of people that we call
{\it collateral sufferers}. This term refers to persons
who are affected by the event under consideration
(whether epidemic, disease or anything else) but do not die.
In order to generate the observed birth dips the size of this group
must be several times larger than the number of fatalities.  

%
\begin{figure}[htb]
\centerline{\psfig{width=8cm,figure=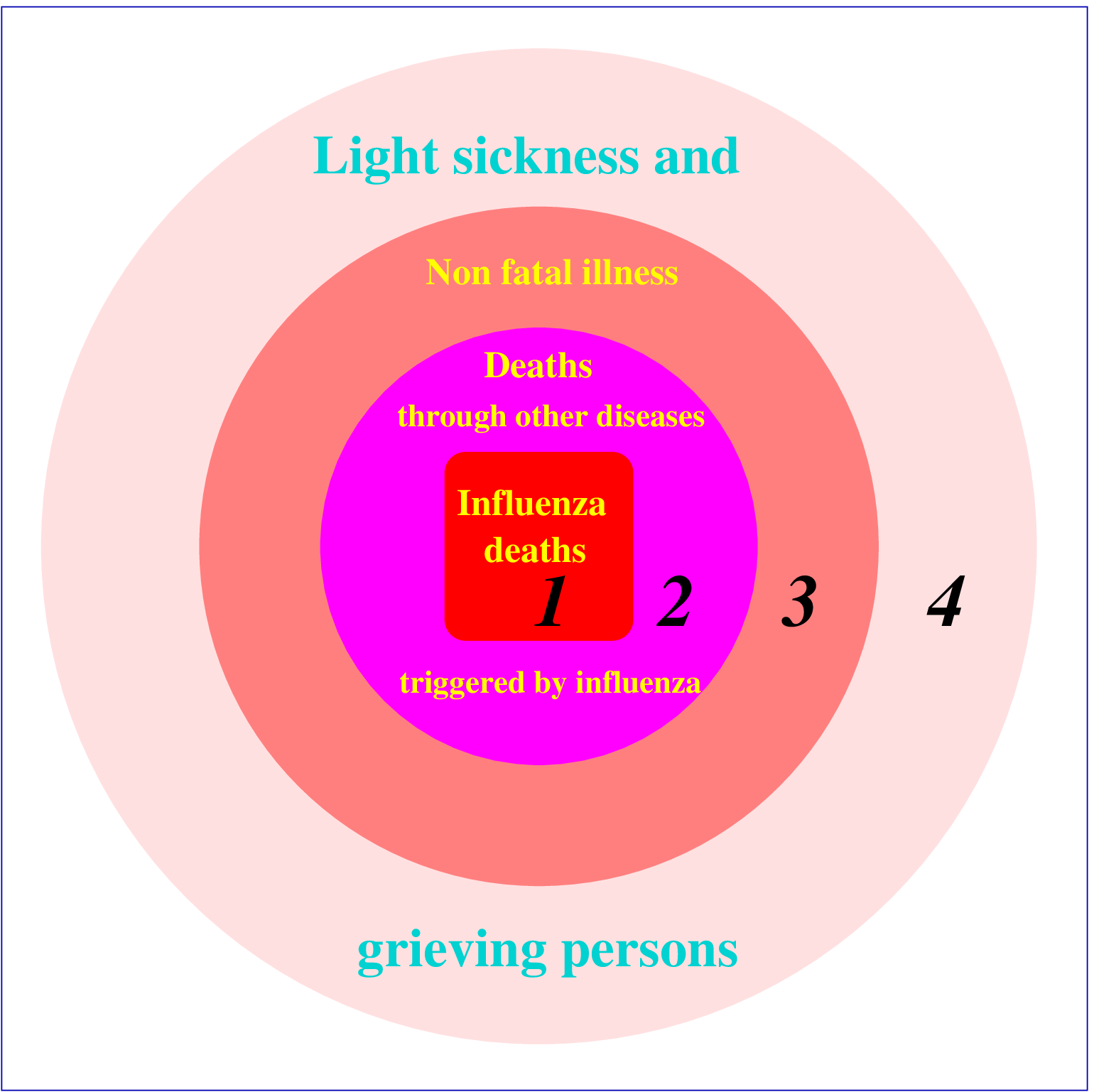}}
\qleg{Fig.\qhu 2b\qhv The impact of a shock ranges from fatalities
to minor afflictions.}
{The figure was drawn for the case of an influenza outbreak but
the same argument applies also to other kinds of shocks.
Thus, for an earthquake some persons are killed,
others only injured and still others 
are unharmed but have their house destroyed.}
{}
\end{figure}
%
This is illustrated in Fig.2b.
In Sweden
although there were only 43,000 excess deaths in 1918,
it is estimated that
about one third of the population, that is to say almost 2 millions
(50 times more than those who died),
was affected by the disease. Naturally, this fraction 
is difficult to define precisely 
because there is a continuum
between the persons who were not affected at all and those who
experienced very mild forms of the disease. 
\qpar

The two following points need to be emphasized.
\qbu Among the persons affected 
by the disease (or even
in the whole population), there may have been a 
more restrained sexual behavior for instance 
by fear of a possible contagion. This 
may have produced 
a reduction in the number of conceptions.
\qbu Among the persons mildly affected by the disease
there may have been biological effects leading
to less fecundation or 
early miscarriages in the 2 or 3 first months of pregnancy.
The fact that in what follows
we will observe the Bertillon
birth effect not only for diseases but also for
earthquakes 
may seem to speak against such biological effects.
However, one cannot exclude possible psychosomatic effects
resulting from a stressful situation. An example is the
so-called famine amenorrhea (Meuvret 1946,
Ladurie 1969, 1975, 1978). 

\qA{Impact of induced delivery}

Finally it should be mentioned that the birth phase 
represented in Fig. 2a
comprises in fact two cases: natural birth and medically assisted
birth through caesarean or induced delivery. Such medical
interventions are much more frequent on ordinary working days than
on holidays. For that reason, there is a drastic 
reduction in daily birth numbers on Saturdays, Sundays and
on holidays. For instance in the United States; 1 January.
4 July, Labor Day, Thanksgiving, 24-25 December
are marked by a reduction of about 30\%.
This effect is of
great importance when one uses daily data; at the level of 
monthly data the effect is ``diluted'' because the deliveries
are simply postponed by a few days, thus the total monthly
figure should not be affected.

\qA{Marriage date does not play a great role}

At first sight one might think that birth dates may be critically
affected by marriage dates. However, in the discussion of
Appendix B it is shown that this effect is fairly weak.
One reason for that is because only first born children are
concerned. Thus, in the 19th or early 20th
century when having four or five
children was common, only a small fraction of the births were
affected. 

\qA{Length of time between sexual intercourse and birth}

It is currently said that, as indicated in Fig. 2a,
pregnancy lasts 9 months which corresponds to $ 9\times 365/12=274 $
average days.
However this statement raises two questions.
\qee{1} For this 9-month estimate what starting point is considered?
\qee{2} What is the dispersion of pregnancy around its average?
This question is of importance because it affects the  width
of the birth trough.
\qpar

These questions are discussed in Appendix C; it appears that 
the average time interval between sexual intercourse and birth can
be taken equal to:\qL
\centerline{ $ 267\pm \sigma $ days, where $ \sigma=9 $ days.}

The fact that the standard deviation is of the order of a few days
shows that the dispersion of the births (that is to say the width
of the dip) will be mostly due to the dispersion of the conceptions.
Even for a sharply defined event such as an earthquake, the 
behavior of collateral sufferers may be modified for several weeks.
Actually, it is through the width of the dip that we can know
how the survivors reacted to the shock in terms of intercourse
frequency.

\qI{Case-study 1: The Great Famine in Finland (1868)}

\qA{How did it happen?}

In Finland the Spring of 1867 was unusually cold; in May 1867
the average temperature was only 1.8 degree Celsius, which is some 8
degrees below the long-term average.  This made it very difficult
to grow spring cereals or potatoes. Then, in autumn the winter started
early and was also colder than average. At the end of 1867 and 
early 1868 relief grains were
imported thanks to a loan of the Rothschild Bank in Frankfurt,
however, as often in such cases, the
inadequacy of the transportation network hampered delivery%
\qfoot{In Sweden where there was a similar yet less serious
situation. According to contemporary newspaper
sources, it seems
that a large part of the relief ended in the pockets
of officials instead of reaching those really in need
(see the Wikipedia article entitled ``Famine of 1866--68'').}%
.
\qpar
In 1867 the death rate
was 38 per thousand, already 35\% above average; then in 1868
it climbed to 78 per thousand. Just as an element of comparison,
it can be mentioned that this rate was 2.8 times higher than
the rate of 28 per thousand reached during the famine of 1961 in China.
In the breakdown of the deaths according to their causes
only 2,350 famine deaths were recorded whereas 27,215 deaths
were attributed to tuberculosis, dysentery, smallpox and whooping
cough. An additional 59,717 were attributed to ``various fevers''
(Finland 1902, p. 416-417). These numbers illustrate the zones
of Fig. 2b; food scarcity was the triggering factor but it
killed very few people directly.
\qpar
With respect to a population of 1.7 million,
the excess-deaths of 80,000 correspond to a rate of 47 per thousand.

\qA{Test of the Bertillon birth effect}

%
\begin{figure}[htb]
\centerline{\psfig{width=17cm,figure=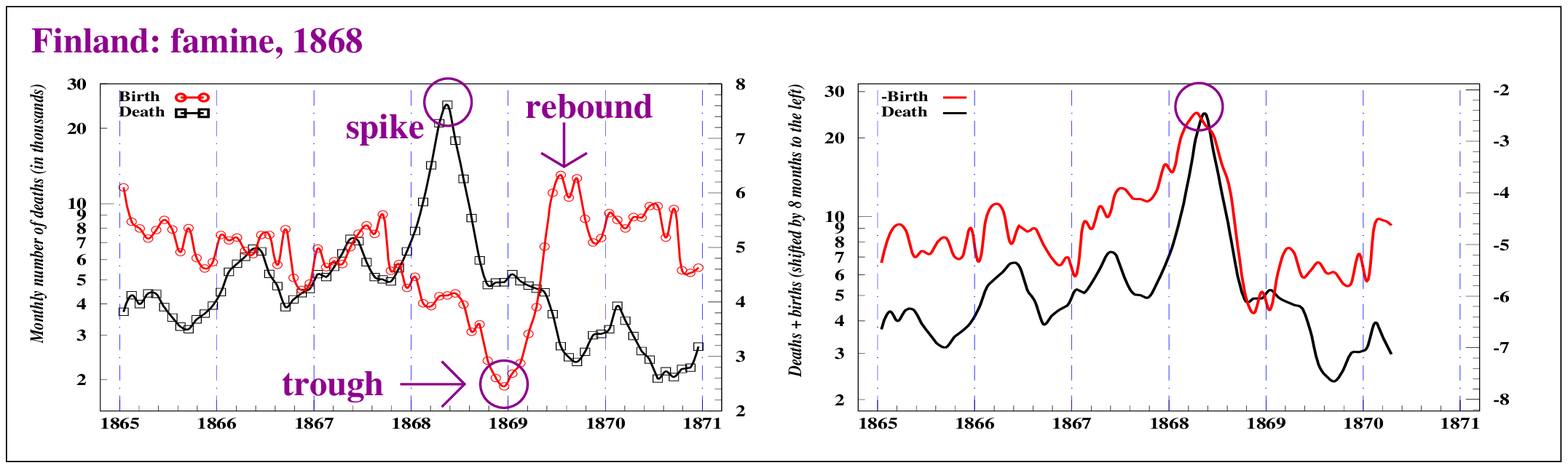}}
\qleg{Fig.\qhu 3a,b\qhv Deaths and births in Finland
during the famine of 1867-1868.}
{Fig. 3a shows the two curves in normal 
graphical representation;
the death scale is on the left and the birth scale on the
right (both are expressed in thousands). 
In Fig. 3b the birth curve was shifted
8 months to the left and turned upside down
(by turning to negative numbers). 
As a result this new
series represents the conceptions (at this point we do not know why
the time lag is rather 8 months instead of 9 months).
The Bertillon birth effect is clearly visible in the 
fact that the maximum of the deaths coincides with a minimum
of the conceptions. In Fig. 3b the correlation of the two series
is $ 0.76 $.
A correction was performed on the monthly
data so as to give all months the same length of 365/12=30.42 days.
All monthly data used in the rest of the paper were corrected in this way
before being used.}
{Source: Finland 1902.}
\end{figure}

Of the mortality cases that we are going to examine, 
this one is the most massive. Thus, one expects the Bertillon
birth effect to be clearly visible and indeed it is. 
For the respective peaks the birth-death 
ratio is approximately 40\%/500\%=0.08.
In other words, if one considers the mortality increase as
the signal and the change in births as the response of the system
we have here an output signal which is 
about 1/10 of the input signal.
\qpar

Incidentally, it can be noted that these data were available in 
Bertillon's time but he did not use them.

\qI{Case-study 2: The influenza pandemic of 1889-1890}

\qA{Circumstances}

The pandemic of December 1889 -- January 1890 is described
in Bertillon (1890). In Paris the first cases occurred in
the week of 17 November 1889. On the basis of death certificates
in Paris
influenza was given as the direct cause of death for only
250 persons. However, during the time of the epidemic there
were about 5,000 more deaths than in the corresponding period
of the previous years. Thus, zone 2 of Fig. 2b was 20 larger
than zone 1. 
With the population of Paris numbering
about 3 million, one gets a fatality rate of 1.6 per thousand.\qL
Whether the epidemic can be called a pandemic is a matter of 
definition. It is true that there was a death spike in many places
(Saint Petersburg, Berlin, Vienna, Paris, London, New York) but
with the exception of Paris and London, in all other places
it was not more severe than the common annual winter death spike.
\qpar

Of the 38 pages of Bertillon's report, only two are devoted
to the effect on birth numbers 9 months later. From the
weekly data that he gives for Berlin and Paris (as
well as some other European capitals) one can draw
the following observations.
\qbu In Berlin the peak of the epidemic occurred in the
last week of December 1889 whereas the lowest point of the
trough of birth numbers 
(with respect to the same weeks of the previous 4 years) 
occurred in the 38th week of 1890.
\qbu In Paris the peak of the epidemic occurred in the
first week of January 1890 whereas the lowest point in the
trough of birth numbers occurred in the 41st week of 1890.
\qL
In both cases the time lag is close to
the 39 weeks of a normal pregnancy.
\qpar

Although Bertillon examines the timing with great accuracy 
he does not discuss the question of the amplitude of the
troughs. In particular, he does not show that the troughs cannot
be explained solely by the deaths due to the epidemic.
That is of course a key point which 
is why in the previous section we discussed it
with great care.

\qA{Test of the Bertillon birth effect}

%
\begin{figure}[htb]
\centerline{\psfig{width=17cm,figure=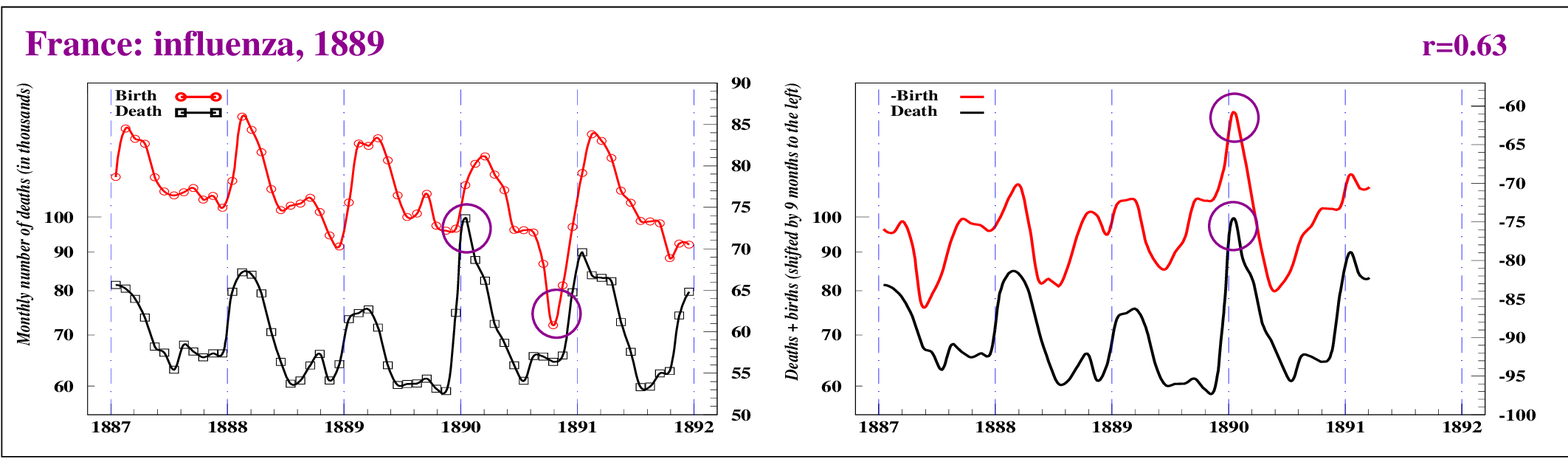}}
\qleg{Fig.\qhu 4a,b\qhv Deaths and births in France
during the influenza epidemic of 1889--1890.}
{In Fig. 4a 
the death scale is on the left and the birth scale on the
right (both are expressed in thousands). 
In Fig. 4b the birth curve was shifted
9 months to the left and turned upside down.
The Bertillon birth effect is again clearly visible, this
time with the expected time lag of 9 months. 
In Fig. 4b the correlation of the two series
is $ 0.63 $.}
{Sources: Statistique de la France, Nouvelle s\'erie. Statistique
annuelle, various years.}
\end{figure}

%
\begin{figure}[htb]
\centerline{\psfig{width=12cm,figure=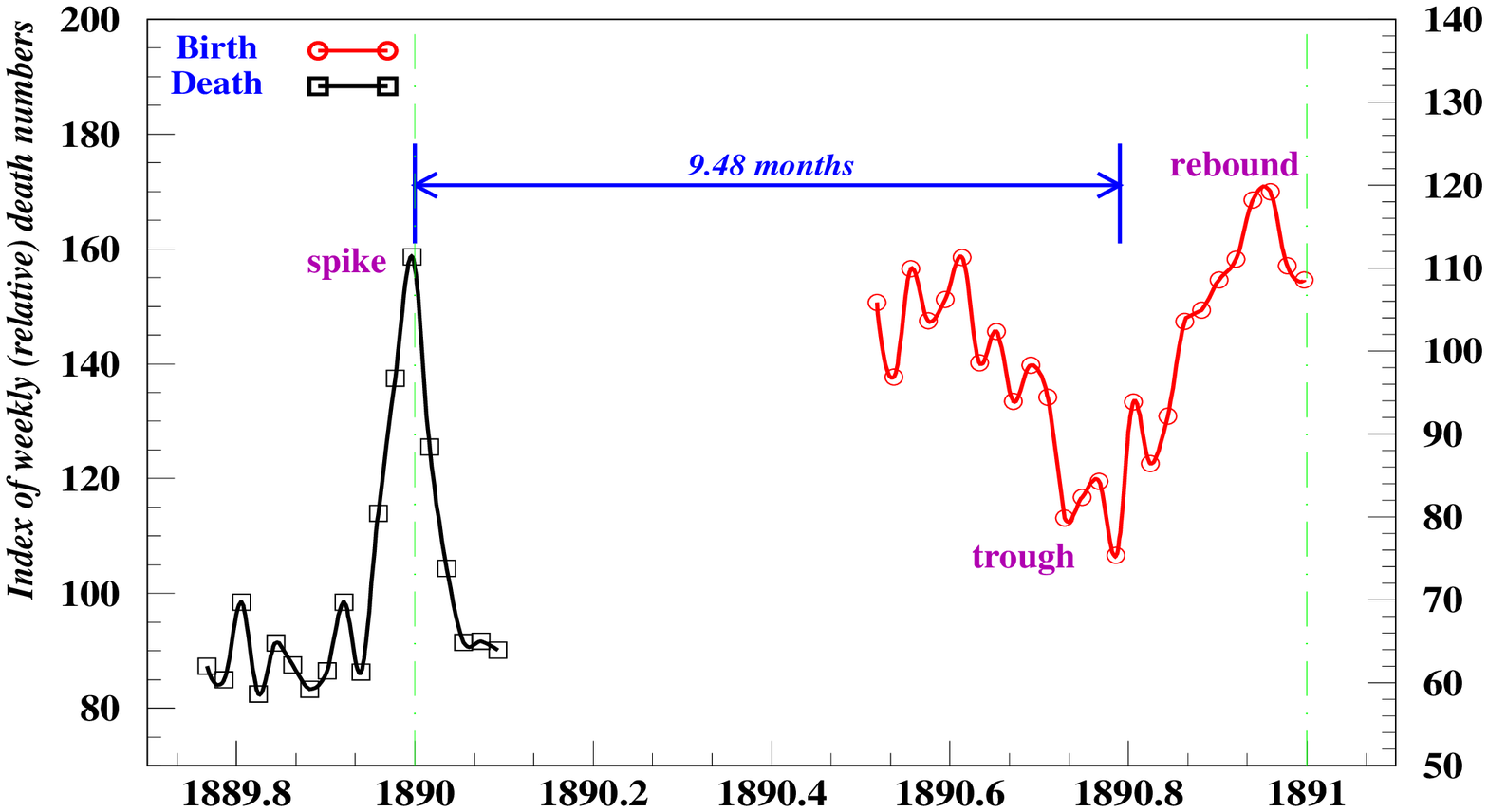}}
\qleg{Fig.\qhu 5\qhv Deaths and births in Berlin
during the influenza epidemic of 1889--1890, weekly data.}
{The death scale is on the left and the birth scale on the
right; the indexes are ratios of the weeks of 1889-1990
to the average of the same weeks in the 4 previous years.
These weekly data provide a more accurate estimate
of the death-birth time lag than the monthly data
used in other graphs.}
{Sources: Bertillon 1892, Statistique G\'en\'erale de la France
1907, p.513.}
\end{figure}

Although the death toll was much smaller than in the case
of Finland, the Bertillon birth effect appears fairly clearly.
What makes the observation more convincing is of course the fact
that the trough can be identified not just in a single
city (where it might occur almost by chance)  
but simultaneously in several cities having non-identical 
seasonal birth fluctuations. In addition to Paris and Berlin,
Bertillon gives also evidence for Barcelona, Rome and Vienna.
In Barcelona and Vienna the lowest point of the birth trough
occurred in September 1890 whereas in Rome it occurred 
in the 3rd week of October 1890.
\qpar

Twenty eight years 
after the pandemic of 1889-1890 there was the great influenza
pandemic of 1918. The death-birth coupling can be observed
in all countries where this disease had a substantial impact%
\qfoot{It can be observed that in the southern hemisphere
(Australia, Chile) it had a different timing than in the
northern hemisphere: instead of October 1918 it peaked
in April 1919. The reason of this lag remains an open
question.}%
. 
However, we will not examine these cases here because they
will be studied closely in the second one of this series of
two papers.

\qI{Case-study 3: The Kanto earthquake of 1923 in Japan}

The particular interest of this case comes from the fact that
it was neither due to famine nor to a disease. 
Although called the ``Great Kanto Earthquake'' in Japan,
it concerned in fact mainly two of the 7 prefectures
which constitute the Kanto Region, namely Tokyo
and Kanagawa (i.e. Yokohama just south of Tokyo).
Some 84\% of the fatalities were concentrated in these
two prefectures.

\qA{Circumstances}

The number of fatalities can be estimated by subtracting the
average death numbers of 1922 and 1924 from the death number
of 1923%
\qfoot{According to the Wikipedia article entitled 
``1923 Great Kanto Earthquake'' there were between 105,000 and
143,000 deaths; this estimate appears to be highly 
exaggerated.}%
, 
i.e. (in thousands): $ 1332 - 1270 = 62 $ (Bunle 1954,
p. 441). For the whole of Japan the death rate was 1.1 per
thousand, in Tokyo it was 8.0 per thousand and in Yokohama
it was 13 per thousand. The earthquake was accompanied by 
a tsunami with a wave up to 13 meter high.
and, especially in Tokyo, by fire tornadoes.

\qA{Evidence}
%
\begin{figure}[htb]
\centerline{\psfig{width=17cm,figure=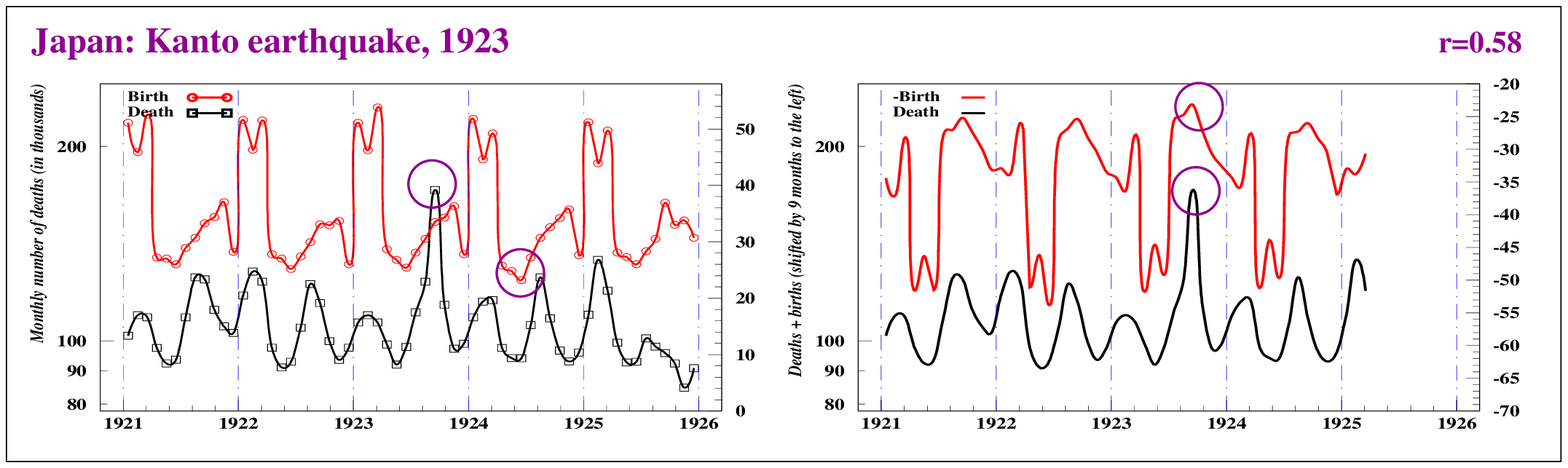}}
\qleg{Fig.\qhu 6a,b\qhv Deaths and births in Japan 
as a result of the Kanto earthquake.}
{In Fig. 6a 
the death scale is on the left and the birth scale on the
right.
In Fig. 6b the birth curve was shifted
9 months to the left and turned upside down.
It is the fact that the seasonal birth fluctuations 
repeat themselves
with great regularity that makes the Bertillon effect
detectable.
As the earthquake concerned only the Kanto region one 
would expect a greater birth effect in the
Tokyo and Kanagawa prefectures.}
{Sources: Deaths: Bunle 1954, p.441; birth rates: Ministry of
Health, Labor and Welfare on the ``Portal site
of Official Statistics of Japan''.}
\end{figure}

The birth trough is of smaller amplitude than in previous cases
and in fact there are
two circumstances which play a crucial role in its identification.
\qee{1} Although the seasonal births have wide
fluctuations (in fact much larger than in other countries)
their annual repetitions are very regular.
\qee{2} As one knows exactly where the trough is expected
even a small signal can be identified.
\qL
A purely statistical analysis that would fail to take into
account these circumstances would result in overestimating the size
of the confidence interval. Incidentally, if monthly data
were available at prefecture level one could get a better
accuracy.

\qI{Case-study 4. The shock of 9/11 in New York City (2001)}

\qA{Circumstances}

Two airliners belonging respectively to United Airlines and
American Airlines
were crashed into the North and South towers of the
World Trade Center complex in New York City%
\qfoot{There have been many odd stories 
and speculations about this attack. 
However, there is one well documented, indisputable
and nevertheless not often mentioned aspect
which
is the huge amount of put options bought in the days
{\it preceding} the attack. For stock owners, put options 
provide a protection against the fall of a stock price because
it gives them the right to sell their stock at a predetermined
price which may be much higher than the current price.
Moreover, because the price of a put option increases when the 
price of the stock declines, speculators can make a profit
by selling their put options. More details can be found
on the following webpage: \qL
http://911research.wtc7.netsept11/stockputs.html.}%
.

%
\begin{figure}[htb]
\centerline{\psfig{width=12cm,figure=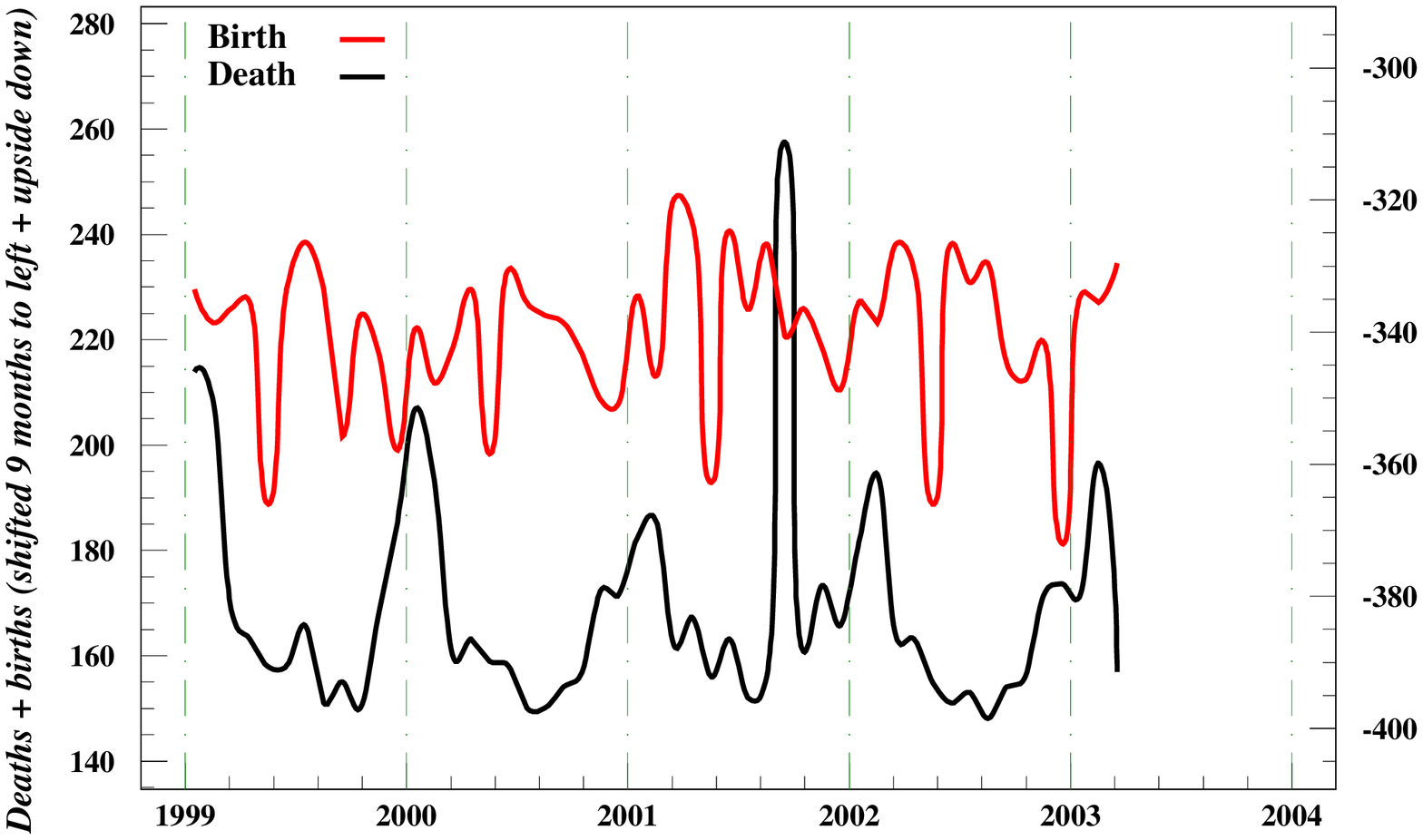}}
\qleg{Fig.\qhu 7\qhv Monthly death and birth
numbers in New York City, 1999-2003.}
{After the birth series has been shifted 9 months to the 
left and reversed one would expect to see 
a peak in September 2001 but there is none.}
{Sources: Summary of vital statistics. The city of New York.
Separate volumes for the years from 1999 to 2003 
(available on Internet).}
\end{figure}
Within about two hours
both buildings collapsed. 
There were some 3,000 fatalities including some 400 firefighters
and police officers. With respect to the 8 million population
of New York city this corresponds to a rate of 0.37 per thousand.

\qA{Evidence}

Fig. 7 is remarkable because it shows that 9/11 did
not produce any birth trough whatsoever. 
It shows that the death-birth coupling
is by no means commonplace.   
\qpar

Here the zones 1 and 2 of Fig. 2b can be merged into one which
corresponds to the total death toll. Zone 3 would correspond
to persons (some 6,000) injured but not killed.
Zone 4 can be seen as 
comprising the families and close relatives  of the persons
killed or injured. Based on the average size of US households
which is of the order of 2.6, one gets for zone 4 a total
number of: $ 2.6\times 9,000=23,400 $. Of this
number only the fraction in the age interval 20-35 would
contribute to the birth trough. Based on the US population
pyramid this fraction is of the order of 15\%.
\qpar
For the pandemics considered so far, it is almost
impossible to estimate the size of zone 4.
However,
for an earthquake one can posit that zone 4 corresponds to the
persons whose houses have been destroyed or damaged.

\qI{Case-study 5. SARS outbreak in Hong Kong (spring 2003)}

The outbreak of SARS (Severe Acute Respiratory Syndrome) 
in Hong Kong is interesting for two reasons.
\qee{1} The number of deaths was small, namely only 300.
With respect to the 6.7 million population of Hong Kong
this represents a rate of only 0.045 per thousand.
Actually the death spike of January-April
2003  was lower than the annual winter 
death spikes of 2004 and 2005.
\qee{2} In 2002, 2004 and 2005 the conception curves
(that is to say the shifted but not reversed birth curve)
has a peak in January and a
trough in mid-August. Yet, in 2003 there is a deep trough in
late March that is to say in coincidence with the height of the
outbreak as defined by the histogram of new cases which shows
a sharp peak in the fortnight 22 March -- 5 April. 

\qA{Circumstances surrounding the death spike}

 It is not only the number of deaths which was small but
also the number of cases, namely 1,730 (i.e. 260 per million
population). Worldwide it was the same picture; there were only
some 700 deaths compared with the 500,000 who died from
influenza in the same year.
Nonetheless, the city took drastic measures.
\qbu Primary and secondary schools were shut for a month beginning in late
 March.
\qbu Various public places were closed.
\qbu Financial companies asked their employees not to come
to their office and work from home.
\qbu  A whole residential complex called Amoy Gardens was put under
an emergency quarantine. The residents were sent to a vacancy center
and the building was closed. Altogether in this complex 329 people
were infected and 42 died. 
\qbu Unlike the influenza pandemic of 1918, SARS was particularly severe
for elderly people.
None of the infected females under 30 died whereas
among males older than 70 the death rate was 75\%.
Thus, although healthcare workers accounted for 23\% of all
infected persons (Leung et al. 2009, p. 14) only few died.

\qA{Statistical evidence}

%
\begin{figure}[htb]
\centerline{\psfig{width=13cm,figure=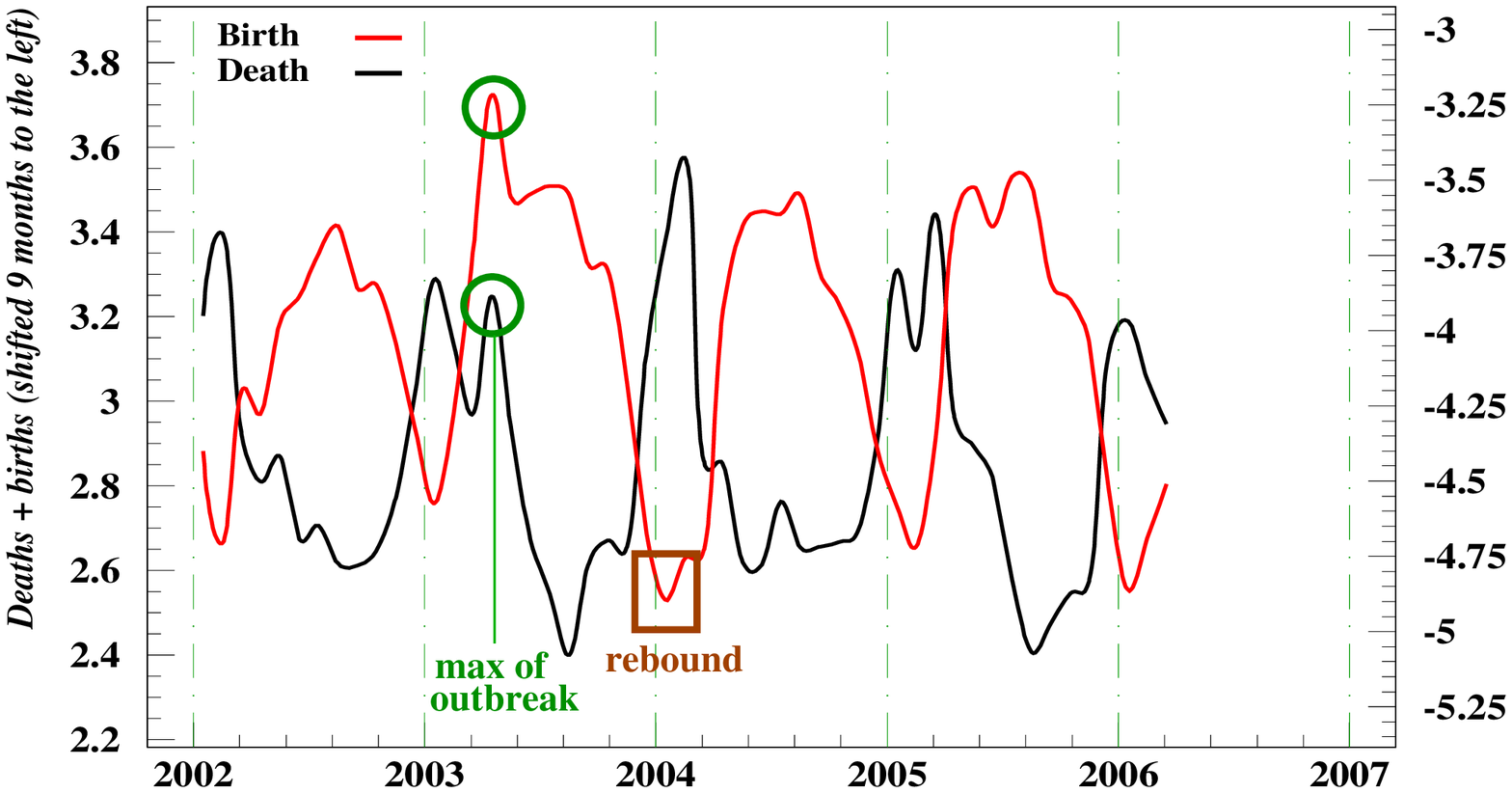}}
\qleg{Fig.\qhu 8\qhv Death spike and birth
trough in Hong Kong
as a result of the SARS epidemic of March--May 2003.}
{Despite the small magnitude of the death spike, 
the trough of the births
is clearly identifiable. However, unlike most other 
previous cases, the correlation between the two curves
is negative (equal $-0.53 $) instead of being positive.
We have here an interesting case where the seasonal winter
death spikes, despite being higher than the SARS spring spike,
do not give rise to any birth trough.}
{Source: United Nations Statistics Division, Demographic
Statistics: Deaths by month of death + Births by month of birth
(available on Internet)}
\end{figure}

Despite the small death toll, 
the death effect of the SARS epidemic can be identified fairly clearly
because it occurs two months later than the standard winter death
outbreaks (Fig. 8). The conception effect can also be identified clearly
the trough occurs of shifted births occurs in March rather than in
August.

\qI{Case-study 6. Earthquake of March 2011 in Japan}

\qA{Circumstances}

The epicenter of the earthquake was under sea near the
city of Sendai which is 300 km north-east of Tokyo. 
The death toll (including the missing) was about 18,400
and a further 6,000 were injured. 
moreover some 400,000 buildings collapsed or half-collapsed%
\qfoot{The source is a report of the Japanese police
published on 10 March 2015 and described in the Wikipedia
article entitled ``2011 Tohoku earthquake and tsunami''
(the Tohoku region is the northern part of the main island
of Japan).}%
.
In other words we can take $ H=400,000 $ as an estimate
of the number of people who were {\it directly} 
affected. We will see below that under appropriate
assumptions one can use this estimate to derive
the expected birth number reduction.

\qA{Statistical evidence}

%
\begin{figure}[htb]
\centerline{\psfig{width=13cm,figure=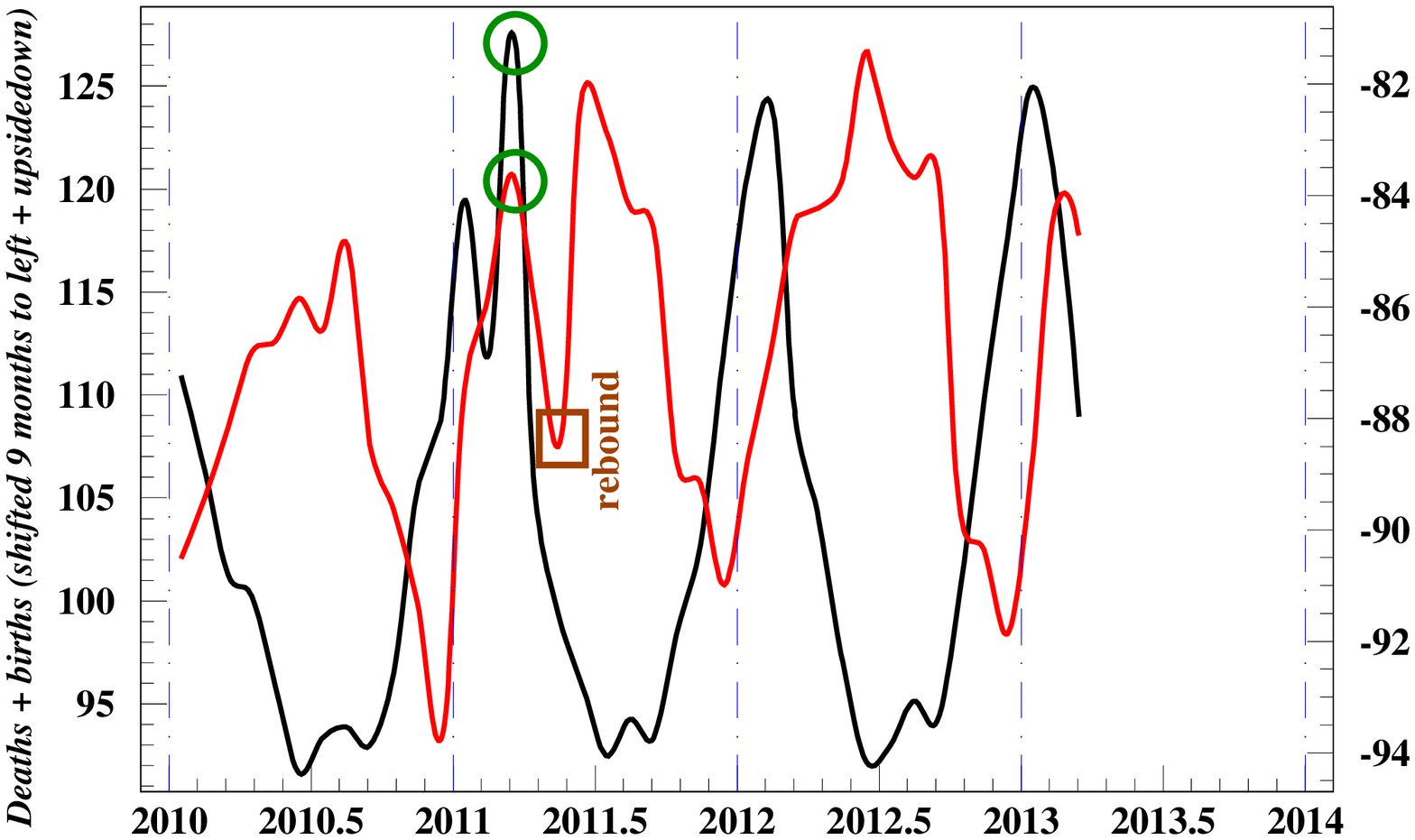}}
\qleg{Fig.\qhu 9\qhv Death spike and birth
trough in Japan
as a result of the earthquake of 11 March 2011.}
{Although the conception trough of March 2011 is not the lowest
it is clearly distinct from the seasonal troughs which occur
in May-July; the same observation applies to the rebound.}
{Source: United Nations Statistics Division, Demographic
Statistics: Deaths by month of death + Births by month of birth
(available on Internet)}
\end{figure}

The conception trough of March 2011 due to the earthquake 
appears fairly clearly
because it is distinct from the seasonal troughs
seen in other years.
\qpar

In 2011 the Japanese crude birth rate was $ \lambda = 8.3 $ 
per 1,000 population.
Let us assume that for the 400,000 persons directly affected
their conception rate in March 2011 was reduced to zero. Thus,
nine months later the birth rate of this group of persons would
also be zero. With respect to a normal year this would result
in a reduction of $ 400,000\times 0.8/1000=3,200 $ births.
Is this number compatible with what is observed?
\qpar

In December 2011 there were $ b_0=83,600 $ births, in December
2010 and December 2012 there were $ b_{-1}=88,100,\ b_1=84,400 $
respectively. Thus the reduction is: $ b_0-(b_{-1}+b_1)/2=2,650 $
which is of an order of magnitude compatible with the 
expected estimate of $ 3,200 $.
\qpar

In all epidemics and disasters, apart from the fatalities,
there is a group of persons which is {\it directly} affected.
For an influenza epidemic this would include the persons
who fell seriously ill but did not die. For an attack like 
9/11 it would be the family members of those who died or
were injured. 
\qpar

Naturally, the previous dichotomous picture
is a simplification. Actually, there
is a gradual transition from those highly affected to 
those lightly affected. Under such an assumption the
reduction $ \Delta B $ in the number of conceptions would be written:
 $$ \Delta B =\int_0^{\lambda} \left(\lambda -x\right)h(x) dx \qn{1} $$
Here $ \lambda $ is the ``normal'' conception rate, $ x $ is the
zone-dependent conception rate and
$ h(x)dx $ is the
number of persons having a conception rate comprised between
$ x $ and $ x+dx $. The dichotomous argument which gives
$ \Delta B=\lambda H $
would
correspond to: $ h(x)=2H\delta(x) $ where $ \delta(x) $ denotes
the Dirac delta distribution and $ H $ the total number of 
people affected.
\qpar

The function $ h(x) $ describes the severity of the shock
among the surviving population. If $ h(x) $ is concentrated
in a narrow interval $ (\lambda -\epsilon,\lambda ) $ it
means that the surviving population is not much affected.
On the contrary, a function $ h(x) $ 
concentrated in a narrow interval $ (0,\epsilon ) $ means
a severe incidence among those affected (their reproduction
rate falls from $ \lambda $ to almost zero.
\qpar

In the next section we discuss shortly how information
about the function $ h(x) $ can be derived from observation.

\qI{Comparison}

\qA{Overview of excess-death rates}

Table 1 summarizes the death rates of the various case
studies considered above.

%
\begin{table}[htb]

\small

\centerline{\bf Table 1 \quad Excess death rates in previous case-studies}

\vskip 5mm
\hrule
\vskip 0.7mm
\hrule
\vskip 2mm

$$ \matrix{
\hbox{Area} \hfill & \hbox{Finland} \hfill & \hbox{Tokyo}\hfill &
\hbox{Massachusetts} \hfill & \hbox{Paris}\hfill & \hbox{New York}\hfill &
\hbox{Hong Kong}\cr
\qtb
\hbox{Year} \hfill & 1868  & 1923  & 1918& 1889 & 2001 & 2003\cr
\noalign{\hrule}
\qth 
\hbox{Death rate} \hfill & 47  & 8.0 &6.4 &1.6  &0.37  &0.045 \cr
\hbox{(per 1,000 population)} \hfill &   &  & &  &  & \cr
\qtb
\hbox{Death-birth coupling} \hfill & \hbox{yes}  & \hbox{yes}  
& \hbox{yes}&  \hbox{yes}&  \hbox{no}& \hbox{yes}\cr
\noalign{\hrule}
} $$
\vskip 1.5mm
Notes: The excess death rates $ d_e $ are defined in the following
way: 
$ d_e= $ Excess deaths with respect to normal years divided by
the population. For instance, in the case of Finland there were
80,000 excess deaths for a population of 1.7 million
which gives $ d_e=80/1.7=47.0 $.
The cases are ranked by decreasing death rate.
While episodes involving diseases extend to the whole country,
episodes involving localized incidents 
(e.g. terrorist attacks or earthquakes) extend
to areas whose definition is somewhat arbitrary. For instance,
whereas the attack of  9/11 concerned a very small area we
computed the death rate with respect to New York City but other 
possible areas would have been Manhattan or New York State.
\vskip 2mm
\hrule
\vskip 0.7mm
\hrule
\end{table}
%

\qA{How to get information about $ h(x) $ from observation}

The function $ h(x) $ describes the incidence of the death spike
in terms of suppressed conceptions and suppressed births.
How can one derive information about it from observation?
One can offer the following suggestions.
\qbu First, is suppressed
conceptions identical to suppressed births? Not necessarily.
There can be a biological regulation which leads to early
elimination of embryos which are not in good shape because
they were generated in conditions of illness, scarcity or famine.
This is a medical question for which one should be able to
find reliable information in the medical literature.
\qbu The fraction of the population directly affected
by the mortality spike comprises the following subgroups.
\qee{i} persons killed ($ K $)
\qee{ii} persons injured or who were seriously ill ($ H_2 $)
\qee{iii} persons
whose living conditions were affected in a major way,
for instance because they lost a family member ($ H_{3a} $) or because
their home was destroyed ($ H_{3b} $).
\qpar

So far, we have focused on death spikes. However,
according to the view outlined above there could
also be cases displaying birth troughs without any 
(or at least very few)
fatalities but instead a substantial number of persons 
who become ill, injured or were otherwise severely affected.\qL
As possible example we investigated
West Nile fewer in the United States.
This disease which is due to a virus carried by mosquitoes
peaks in August-September which would give a birth trough
in May. However, we could not identify any significant
birth trough even in the states most affected,
e.g. Texas in 2012. The number of incapacitated persons
may be too low.

\qA{Why is there no birth trough in the case of 9/11?}

There is no birth trough following 9/11
but there is one in the Hong Kong SARS epidemic
whose death rate is about 10 times smaller. This seems fairly
puzzling but the two events were of very different kinds.
Whereas 9/11 was a one-day event, the SARS epidemic lasted
a few months and, 
especially during early times, the spread of the disease
as well as its severity (in terms of number of deaths per cases)
were a matter of uncertainty and concern. Thus, even persons
who did not become ill were affected.

\qI{Conclusion and predictions}

We have shown that sudden death spikes are almost always followed
9 months later by a birth trough. We have also seen that the
spike of 9/11 did not lead to the expected birth reduction.
This observation is particularly intriguing when compared
with the SARS outbreak in Hong Kong in which there
were less deaths than in 9/11 and which was nonetheless followed
by a birth dip.
Although a tentative explanation was proposed it is clear
that in order to get a better understanding it would be useful
to examine other cases in which a death spike is {\it not} followed
by a birth dip.
\qpar

For that purpose the normal procedure is to propose
predictions in the hope that for some of
them the expected dip will not materialize. 
This strategy is very much in line with what was done
in the development of physics. Every time an
expectation happened to be contradicted by observation, this
was an opportunity for new progress. A well-known
case was the non-observation of the aether 
by Michelson and Morley which led to the theory
of relativity.  
\qpar

There are two possible kinds of predictions. 
\qbu The first kind consists in what we may call standard predictions;
they are instances very similar to those already analyzed
but less well known because of smaller amplitude.
The predictions of birth troughs in the Netherlands in 1920-1922
and in Chile in 1923 are
of this kind. Both graphs are shown in Appendix E which is included
in the arXiv version of this paper.
Although in the case of Chile we do not yet know
the reason of the death spike of July-August 1923, 
the birth trough could indeed be observed in the month 
in which it was expected.
\qbu In the second kind of predictions one deliberately considers
a cause of death that has not yet been tested. One case
of that kind are deaths due to heat waves. For instance
in France in August 2003 a heat wave caused
13,700 excess deaths. This was a fairly exceptional case.
Most other heat waves caused only of the order of one or a few
thousands excess deaths. In contrast with diseases,
for heat waves there is no contagion effect; 
in contrast with earthquakes
there are no collateral destructions. However, in contrast
with 9/11, all persons (at least those 
who do not have air conditioning)
are directly affected to some degree. \qL
Are there birth dips
in the wake of heat waves?
Because of the fairly low amplitude of most death spikes
the identification
of the troughs turns out to be more difficult but 
they are nevertheless present (Rey et al. 2007, 
R\'egnier-Loilier 2010).

\qpar

{\bf Acknowledgments} We wish to express our sincere thanks to 
to Ms. Ela Klayman-Cohen of
the Swedish ``Statistical Central Agency'' (Statistika
Central-Byr\aa ns), Ms. Maija Maronen of ``Statistics Finland'',
Mr. Chihiro Omori of the Japanese Ministry of Internal Affairs
for their kind help
in guiding us through the rich datasets of their respective
countries.

\appendix

\qI{Appendix A: Population fluctuations vs. birth fluctuations}

In this appendix we examine the implication of a constant
birth rate on population and birth changes.
\qpar

The monthly birth rate is defined as: $ \lambda=b(t)/P(t) $
where $ b(t) $ is the number of births in month $ t $. 
Now let us
apply changes $ \Delta b $ and $ \Delta P $ to the numerator
and denominator respectively. How these changes must be connected
if $ \lambda $ is to remain constant is shown by the following
calculation.
$$ \lambda ={ b(t)+\Delta b \over P(t)+\Delta P }=
{ b(t) \over P(t) }{ 1+\Delta b/b(t) \over 1+\Delta P/P(t) } $$
$$ \lambda \sim { b(t) \over P(t) }\left[ 1+{ \Delta b \over b(t) } 
- { \Delta P \over P(t) } \right],\quad \lambda \hbox{ constant }
\Longrightarrow 
\Delta b=\left[ { b(t) \over P(t) }\right]\Delta P $$

In the case of a death spike the population change
will be given by the excess-death number $ \Delta P=-[d(t)-b(t)]=-e $ 
(for the sake of simplicity we ignore
the 9-month time-lag between conception and birth).\qL
This leads to the proposition given in the text.
\qpar

As a case in point in order to illustrate the previous argument
with real data
we consider again Sweden during the influenza
epidemic of 1918.

\qA{Example of Sweden in October 1918}

It is by purpose that we selected a country which did not take
part in the First World War so as to avoid any interference.
The data are taken from Bunle 1954 (p. 313, 438)
and Flora et al. 1987 (p. 73).
\qpar
In early 1918 the Swedish population numbered 5.8 millions; on
average 
its annual birth and death rates were 2.0\% and 1.3\% respectively.
\qpar

\count101=0  \ifnum\count101=1
In October 1918 there were 17,278 deaths and 9,853 births
that is to say an excess death of $ d=7,425 =17278-9853 $ (by contrast in
October 1917 the population had increased by $ 4,157 $).
Assuming a constant monthly birth rate $ 2/12=0.16\% $
this excess death would lead to $ \Delta b=7426/600=12 $.
Let us compare this number with the births actually observed in 
July 1919.
\qpar
In July 1919 there were 7,703 live births
instead of an average of 10,200 in the months of July of 1916, 1917
and 1918. Thus, $ \Delta b=2,497 $, which is 208 times more
than the 12 expected under the assumption of a constant birth rate.
\fi

Whereas in ``normal'' years (e.g. in the adjacent years 1917,1919)
there were 5,862 October deaths, in October 1918 there were 17,278
deaths which represents an excess-death number of 11,416.
With a constant monthly birth rate of $ 2/12=0.16\% $ these
excess deaths would result in a birth deficit of 
$ 11,416\times 0.16/100=18 $.  Let us compare this number with the
births actually observed in  July 1919.
\qpar

Whereas in ``normal'' years there were on average 10,618 July
births, in July 1919 there were only 7,703 which represents a
birth deficit of 2,465. This is 136 times more than the 18
expected under the assumption of a constant birth rate.

\qpar
In other words, the birth number reduction cannot simply be a
``mechanical'' consequence of the death spike. It
can only be explained by a drastic reduction in
the birth rate. 

\qI{Appendix B. Weak effect of marriage postponement}

An explanation based on the number of marriages may
be considered but in fact it can be quickly discarded.
One may say that 
during the time of the
epidemic people postponed planned marriages. 
As in Sweden in any normal month
there were about 4,000 marriages
a drastic reduction could in principle 
account nine months later for
a fall in births of the same magnitude, that is to say
of the size actually observed.
However,
it appears that in most countries there was only a slight
reduction in the number of marriages or even none at all.
Thus, in Sweden in October 1918, at the height of the
influenza epidemic,
there were 4,323 marriages
compared to an average of 4,255 in the same month of
October 1915,1916,1917.
\qpar

Even when there is a fall in marriages it has not 
necessarily an impact on the number of births 9
months later. An illustration is provided by New York City
in September 2001. As a result of the World Trade Center attack
the number of marriages fell from 6,753 in August 2001 to 2,616 in
September. However the correlation between the time series
of monthly percentage
variations of marriages and conceptions over the 60
months from 1999 to 2003 is as low as 0.025 (which,
for a confidence level of 0.95, is not a significant correlation).
The corresponding linear regression reads:
$ \Delta b/b=0.003(\Delta m/m)+0.11 $ which shows that the
observed fall of 61\% of marriages
will translate in a fall of less than 1\%
for the births.
\qpar
In short, in some cases where there is both a substantial
reduction in the number of marriages (which is rare)
and a significant
marriage-birth correlation, this factor may
contribute but it
cannot be the root factor that we are looking for.

\qI{Appendix C: Length of time between conception and birth}

The standard length of pregnancy, namely 40 weeks (280 days) is 
counted from the woman's last period, not the date of conception which
generally occurs two weeks later.
For the present study we rather need the time $ D $
between the sexual intercourse
which led to the conception and birth. The conception results
from the encounter between a spermatozoon and an ovum (also called
ovule or egg cell). Ovulation means that the egg is
released from the ovary into
the Fallopian tube; it remains there in good shape for only one day
which means that conception and ovulation must take place almost
on the same day. In contrast the spermatozoa can stay alive in the
Fallopian tube for 3 days with similar conception probabilities
during those 3 days%
\qfoot{Actually they can remain alive for about 6 days but the
probability of conception during these three extra days is only
one third of the conception probability during the first 3 days
(Wilcox et al. 1995).}%
.
Once fertilized the egg starts its journey down the Fallopian tube
and into the uterus where it will get implanted.
\qpar

%
\begin{figure}[htb]
\centerline{\psfig{width=14cm,figure=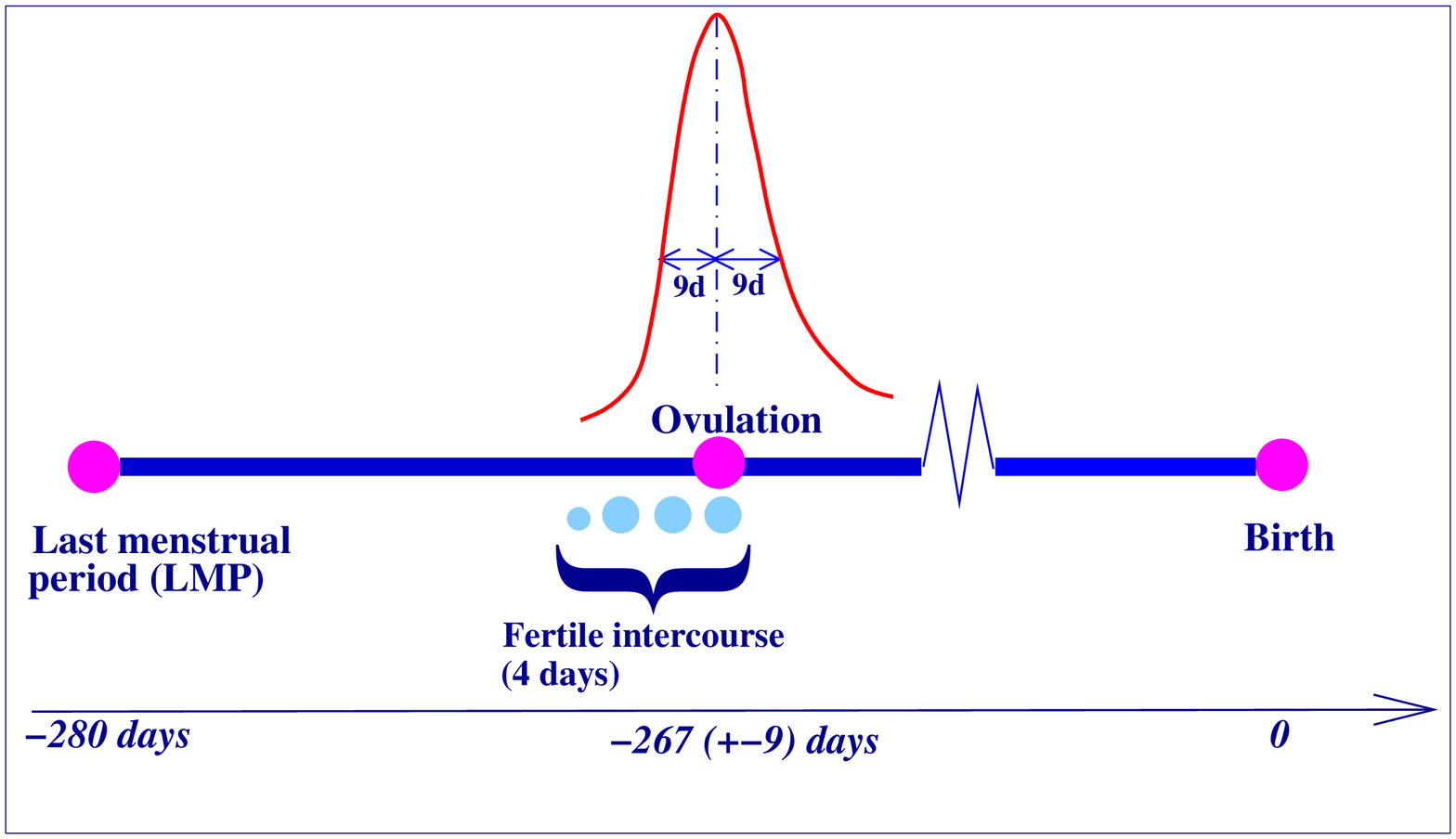}}
\qleg{Fig.\qhu C1\qhv Average time interval between intercourse conducive
to conception and birth.}
{The familiar estimate of 9 (mean) months corresponds to
an interval of 274 days, that is to say one week longer than the
(more accurate) estimate given in the figure. The standard 280 day
figure refers to the time interval between the last menstrual period
and birth; it overestimates the time between sexual intercourse
and birth by two weeks. 
The 9 day dispersion of ovulation refers to
its standard deviation.}
{Sources: Wilcox et al. (1995,2000), Bhat et al. (2006)}
\end{figure}
%

The key-question then is: when does ovulation occur?
It is often said that it occurs at day 14 in the menstrual cycle.
In fact, this depends upon the length of the
menstrual cycle (Wilcox et al. 2000). When the length of the cycle is
28 days, ovulation occurs on average at day 12. When the cycle
lasts more than 30 days, the ovulation takes place on day 14.
The global average for all cycles is 14 days. Coupled
with the standard estimate of 280 days starting from the beginning of
the  cycle one gets: 280-14=266 days following ovulation. 
With respect to intercourse one gets time intervals of
$ D=266, 267, 268 $ days with  same probability for each duration,
which gives an average of $ D=267 $.
The distribution of the fertility window is approximately
Gaussian
\qfoot{Actually with 80\% of the cases within $ \pm \sigma $
(instead of 60\% for a Gaussian)
the distribution is somewhat more narrow than a Gaussian
(Bhat et al (2006)).}
with a standard deviation of 9 days. In summary:\qL
 \centerline{$ D=267\pm \sigma,\ \sigma=9 $ days}
\qpar

This estimate is confirmed by the following result 
based on a sample of 125 pregnancies in which
``the median time from ovulation to birth was 268 days''
(Jukic et al. 2013).
\qpar

{\bf Remark}.\quad In the previous results ovulation
time was determined by urinary hormone measurements
or estimated through ultrasound observation performed
later on in the pregnancy. Needless to say each of these
methods involve some uncertainty.
It might seem that medically assisted
conception would afford a direct and henceforth
more accurate method. The difficulty here is that
such pregnancies are know to lead
to an inflated proportion of preterm deliveries. 

\qI{Appendix E: Test of the Netherlands and Chile predictions}

Can we use what we have learned in this study to make predictions?
Basically, what we have seen is that major mortality surges
result 9 months later in a dip of live births. So far, we have 
restricted ourselves to really big events. Here we will test
predictions based on smaller and less known mortality surges.
\qbu The first test is based on the influenza epidemic
of January 1922. In most countries the outbreak resulted in
only a slight increase of death numbers. For instance, in France there
were 13,000 more deaths than in the same month
of 1921 and 1923, which represents an increase of 21\%.
In Japan
it resulted in a mortality increase of 13\%. 
\qpar

However, there were
two countries which were hit much more severely, namely the
Netherlands and Scotland. In these places the death rate
of January 1922 was twice its value in 1921 or 1923.
It may appear surprising to mention Scotland without
Britain (i.e. England and Wales). It is because for Britain we have
only quarterly data. For the first quarter
these data reveal 39,000 excess deaths and if one assumes
that they were spread over two months (as is
the case in Scotland) one gets severity estimates which are almost as
high as in Scotland.  
\qpar

With a doubling in death rate, there should be a clear-cut 
Bertillon effect in September-October 1922. Is this confirmed
by observation? Fig. 9 shows that there is 
indeed a visible birth trough. In addition, one sees a smaller
trough on the left side of the graph which is due to the small
influenza outbreak of February 1920 (this outbreak
was stronger in the US than in Europe). For 
Scotland (not shown here) the prediction is similarly
confirmed by observation.
\qbu The second example concerns an excess mortality that occurred
in Chile in July-August 1923 and caused about 7,000 excess deaths
over the two months. Actually this death surge remains somewhat mysterious
in the sense that its cause remains unknown.
It was not an influenza epidemic because there
are no excess deaths whatsoever in neighboring countries such as
Argentina or Uruguay. 
The fact that the mortality
extends over 2 months excludes an earthquake and indeed no earthquake
was recorded in July-August (though there was one on 4 May 1923).
Despite of this uncertainty the birth trough is indeed present
where expected.

%
\begin{figure}[htb]
\centerline{\psfig{width=12cm,figure=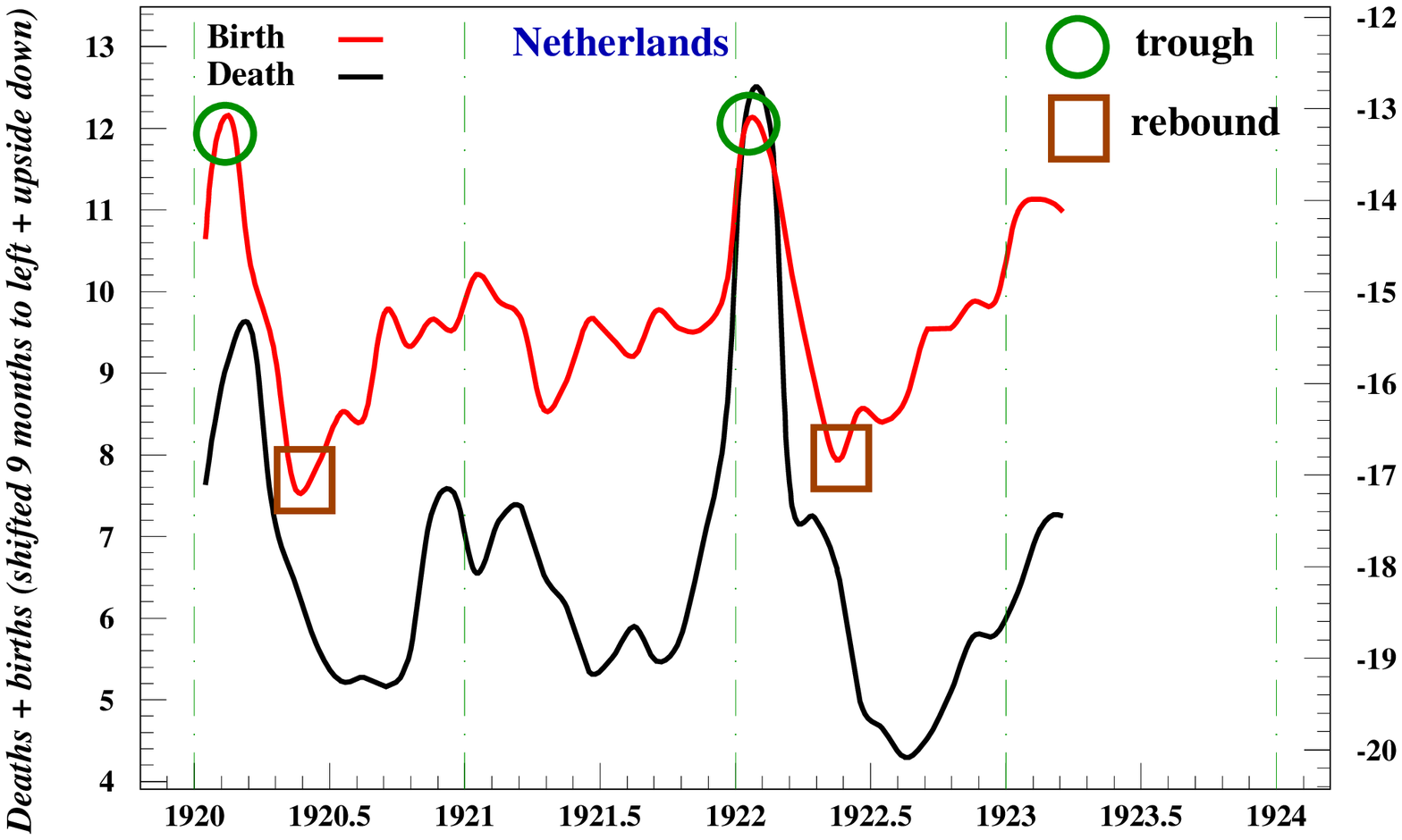}}
\qleg{Fig.\qhu E1\qhv Prediction of birth troughs
in the Netherlands.}
{The influenza epidemic of January 1922 was particularly
severe in the Netherlands. The expected effect was a birth
trough in September-October 1922 followed by a rebound.
Both can indeed be observed.
In addition 
the trough (i.e. the peak in the inverted scale) on the 
left-hand side of the graph is due to the (small) mortality
surge of February 1920 (a late replica of the epidemic of 1918).}
{Source: Bunle (1954)}
\end{figure}
%
\begin{figure}[htb]
\centerline{\psfig{width=12cm,figure=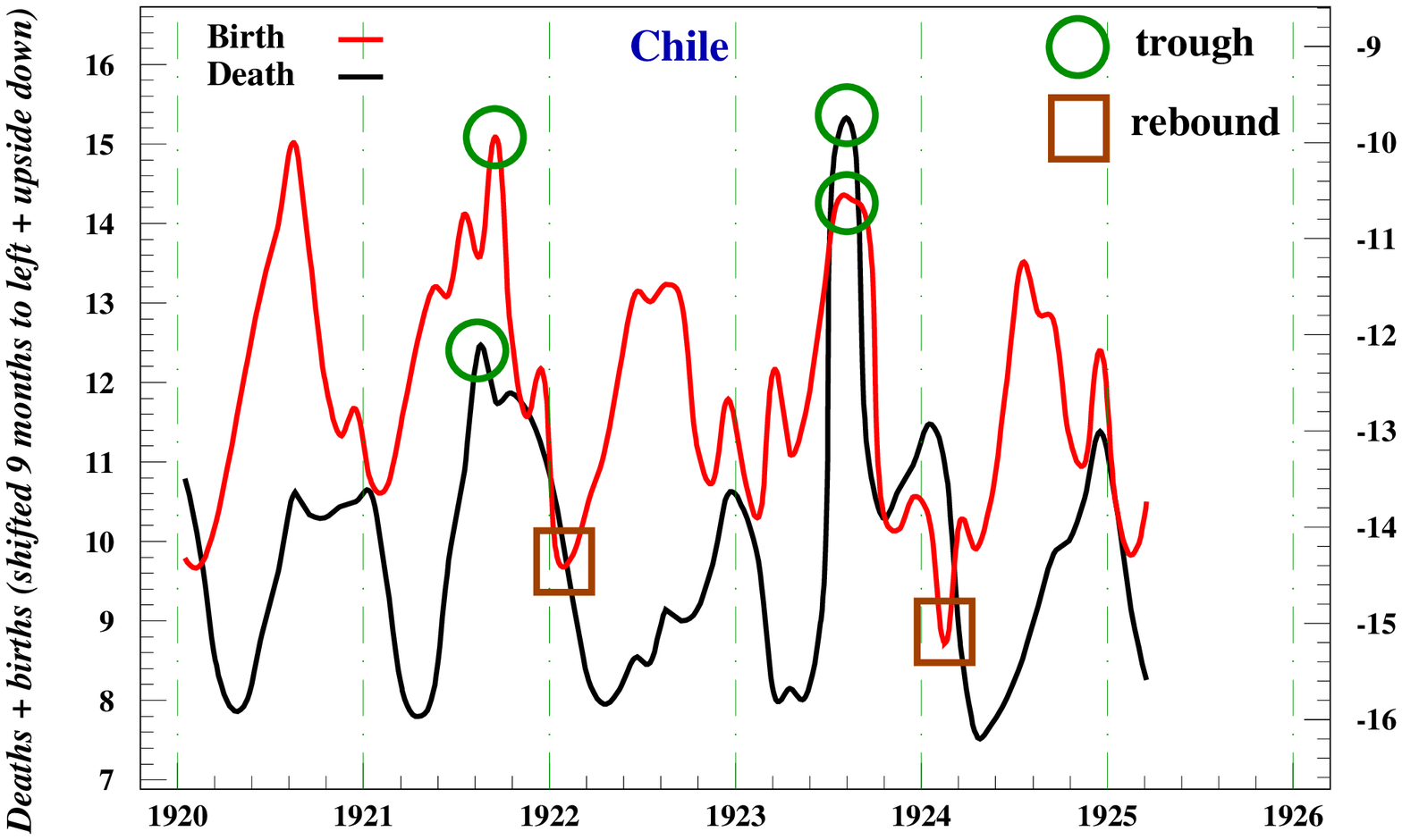}}
\qleg{Fig.\qhu E2\qhv Prediction of birth troughs
in Chili.}
{The mortality surge of July-August 1923 was due to a yet
unknown factor. Nonetheless it resulted in the expected birth
trough. It is the association trough-rebound which allows us to
distinguish the troughs due to a death surge from the seasonal
troughs which also occur in mid-year but have no associated
rebound.}
{Source: Bunle (1954)}
\end{figure}

\vskip 10mm

{\bf References}

\qparr
Bhat (R.A.), Kushtagi (P.) 2006: 
A re-look at the duration of human pregnancy.
Singapore Medical Journal 47,12,1044-1048.

\qparr
Bertillon (J.) 1892: La grippe \`a Paris et dans quelques
autres villes de France et de l'\'etranger en 1889--1890
[The influenza epidemic in Paris and in some other cities
in western Europe].
In : Annuaire statistique de la ville de Paris pour l'année 1890, 
p. 101-132.
Imprimerie Municipale, Paris.\qL
[Available on Internet, for instance at the following address:\qL
{\small 
http://www.biusante.parisdescartes.fr/histoire/medica/resultats/index.php?do=livre\&cote=20955}

\qparr
Bunle (H.) 1954: Le mouvement naturel de la population dans le
monde de 1906 \`a 1936. [Vital statistics of
many countries world-wide from 1906 to 1936.]
Editions de l'Institut d'Etudes D\'emographiques, Paris.

\qparr
Finland 1902. The French subtitle of this official
publication is: \'El\'ements d\'emographiques principaux de la
Finlande pour les ann\'ees 1750-1890, II: Mouvement de la population.
The Finnish title is:  Suomen [Finland] V\"aest\"otilastosta
[demographical elements] vuosilta [years] 1750-1890, II:
V\"aest\"on [population] muutokset [changes]. Helsinki 1902. \qL
[Available on line, 526 p.]

\qparr
Einstein (A.), Podolsky (B.), Rosen (N.) 1935: 
Can quantum-mechanical
description of physical reality be considered complete?
Physical Review 47,10,777–780.

\qparr
Flora (P.), Kraus (F.), Pfenning (W.) 1987: State, economy, and
society in western Europe 1815-1975. A data handbook in two
volumes. Volume 2: The growth of industrial societies and
capitalist economies.
Macmillan Press, London.

\qparr
Garenne (M.) 1994: Sex differences in measles mortality:
a world review.
International Journal of Epidemiology 23,3,632-642.

\qparr
Garenne (M.), Lafon (M.) 1998: Sexist diseases.
Perspectives in Biology and Medicine 41,2,176-189.

\qparr
James (W.H.) 1971: Cycle day of insemination, coital rate and
sex-ratio. Lancet 7690, 112-114.

\qparr (W.H.) 2006: Offspring sex ratios at birth as markers
of paternal endocrine disruption.
Environmental Research 100,1,77-85.

\qparr
James (W.H.) 2009: The variations of human sex ratio at birth during
and after wars, and their potential explanations.
Journal of Theoretical Biology 257,116-123.

\qparr
James (W.H.) 2010: The inconstancy of human sex ratios at birth.
Eletters to the Editor, Fertility and Sterility (website) 27 May
2010. 

\qparr
Jukic (A.M.), Baird (D.D.), Weinberg (C.R.), McConnaughey (D.R.),
Wilcox (A.J.) 2013:
Length of human pregnancy and contributors to its natural variation.
Human Reproduction 28,10,2848–2855.

\qparr
Ladurie (E.L.) 1969: L'am\'enorrh\'ee de famine (XVIIe-XXe si\`ecles)
Annales. \'Economies, Soci\'et\'es, Civilisations 24,6,1589-1601.

\qparr
Ladurie (E.L.) 1975: Famine amenorrhea (seventeenth to twentieth 
century) in Forester (R.) and Ranum (O.) editors: Biology
of men in history. Baltimore.

\qparr
Ladurie (E.L.) 1978: Die Hungeramenorrh\"oe (17.-20. Jahrhundert).
in Arhtur E, Imhof, editor, Biologie des Menschen in der Geschichte.
Beitr\"age zur Socialgeschichte der Neuzeit aus Frankreich
or Skandinavia.  Stuttgart-Bad Cannstaat.

\qparr
Leung (G.M.), Ho (L.M.), Lam (T.H.), Hedley (A.S.) 2009: 
Epidemiology of SARS in the 2003 Hong Kong epidemic.
Hong Kong Medical Journal 15,12-16.

\qparr
Meuvret (J.) 1946: Les crises de subsistence et la d\'emographie
dans la France d'Ancien R\'egime. 
[Food scarcity crises in France before 1789.]
Population 643-650.

\qparr
Mitchell (B.R.) 1978: European historical statistics 1750-1970.
Macmillan, London.

\qparr
Polasek (O.), Kolcic (I.), Rudan (I.) 2005: Sex ratio at birth
and war in Croatia (1991--1995).
Human Reproduction 20,2489-2491. 

\qparr
R\'egnier-Loilier (A.) 2010: \'Evolution de la saisonnalit\'e des
naissances en France de 1975 \`a nos jours [Changes in the seasonal
birth pattern in France from 1975 to 2006].
Population, 65,1,147-189.

\qparr
Rey (G.), Fouillet (A.), Jougla (E.), H\'emon (D.) 2007:
Vagues de chaleur, fluctuations ordinaires des temp\'eratures
et mortalit\'e en France depuis 1971 [Heat waves, temperature
fluctuations and mortality in France from 1971 to 2003].
Population 62,3,533-564.


\qparr
Sardon (J.-P.) 2005: Influence des \'epid\'emies de grippe
sur la f\'econdit\'e. [Influence of influenza epidemics on
fertility.]
In: Bergouignan (C.) et al.: ``La population de la France.
Evolutions d\'emographiques depuis 1946, Vol. 1, p. 413-417.\qL
[This short paper is focused
on the effect of the epidemic of 1957 on births in European countries.
Unfortunately the ``taux conjoncturel de f\'econdit\'e''
that is represented in the graphs
is a (virtual) previsional rate rather than
an actual rate.]

\qparr
Statistique G\'en\'erale de la France 1907: Statistique internationale
du mouvement de la population d'apr\`es les registres 
d'\'Etat Civil. R\'esum\'e r\'etrospectif depuis l'origine des
statistiques de l'\'Etat Civil jusqu'en 1905. 
[International vital statistics based on official records.
Historical summary from the start of official statistics to 1905.]
Imprimerie Nationale, Paris.

\qparr
Sweden 1863. The Swedish title of this official publication is:\qL
Befolknings-Statisk [population statistics] II. 1.
Underd\aa niga Ber\"attelse [report to the King] 
f\"or \aa ren 1856 med 1860 [for the years 1856-1869]
Statistika Central-Byr\aa ns [Statistical Central Agency].
Stockholm, 1863. 

\qparr
Wilcox (A.J.), Weinberg (C.R.), Baird (D.D.) 1995: 
Timing of sexual intercourse in relation to ovulation. Effects on the
probability of conception, survival of the pregnancy, and sex of the
baby.
New England Journal of Medicine 333,1517-1521 (7 December 1995).

\qparr
Wilcox (A.J.), Dunson (D.), Baird (D.D.) 2000:
The timing of the ``fertile window'' in the menstrual cycle: day
specific estimates from a prospective study.
British Medical Journal 321, 7271, 1259–1262 (18 November 2000).

\end{document}